\documentclass[letterpaper,twocolumn,10pt]{article}
\usepackage{usenix-2020-09}

\usepackage{tikz}
\usepackage{amsmath}
\usepackage{amssymb}
\usepackage{multirow}
\usepackage{booktabs}
\usepackage{pifont}
\usepackage{makecell}
\usepackage{algorithm}
\usepackage{algorithmic}
\usepackage{hyperref}
\usepackage{dsfont}
\usepackage{balance} 
\usepackage{xcolor}
\usepackage{ulem}
\usepackage{authblk}
\usepackage{marvosym}

\newcommand{\partitle}[1]{ \noindent \textbf{#1.}}

\begin{document}

\date{}

\title{\Large \bf Membership Inference Attacks Against Vision-Language Models}


\author{
{\rm Yuke Hu\thanks{Work done during a visit at CISPA\\ \indent\ \Letter\ Corresponding Author} \textsuperscript{1}}
{\rm Zheng Li\textsuperscript{2}}
{\rm Zhihao Liu\textsuperscript{1}}
{\rm Yang Zhang\textsuperscript{3}}
{\rm Zhan Qin\textsuperscript{\Letter1}}
{\rm Kui Ren\textsuperscript{1}}
{\rm Chun Chen\textsuperscript{1}}
}
\affil{
{\rm \textsuperscript{1}The State Key Laboratory of Blockchain and Data Security, Zhejiang University\\}
{\rm \textsuperscript{2}Shandong University  }
{\rm \textsuperscript{3}CISPA Helmholtz Center for Information Security}
}

\maketitle

\begin{abstract}
Vision-Language Models (VLMs), built on pre-trained vision encoders and large language models (LLMs), have shown exceptional multi-modal understanding and dialog capabilities, positioning them as catalysts for the next technological revolution. 
However, while most VLM research focuses on enhancing multi-modal interaction, the risks of data misuse and leakage have been largely unexplored.
This prompts the need for a comprehensive investigation of such risks in VLMs.

In this paper, we conduct the first analysis of misuse and leakage detection in VLMs through the lens of membership inference attack (MIA).
In specific, we focus on the instruction tuning data of VLMs, which is more likely to contain sensitive or unauthorized information. 
To address the limitation of existing MIA methods, we introduce a novel approach that infers membership based on a set of samples and their sensitivity to temperature, a unique parameter in VLMs. 
Based on this, we propose four membership inference methods, each tailored to different levels of background knowledge, ultimately arriving at the most challenging scenario. 
Our comprehensive evaluations show that these methods can accurately determine membership status, e.g., achieving an AUC greater than 0.8 targeting a small set consisting of only 5 samples on LLaVA.\footnote{Code is available at \url{https://github.com/YukeHu/vlm\_mia}.}

\end{abstract}

\section{Introduction}
\begin{figure}[t]
    \centering
    \includegraphics[width=0.45\textwidth]{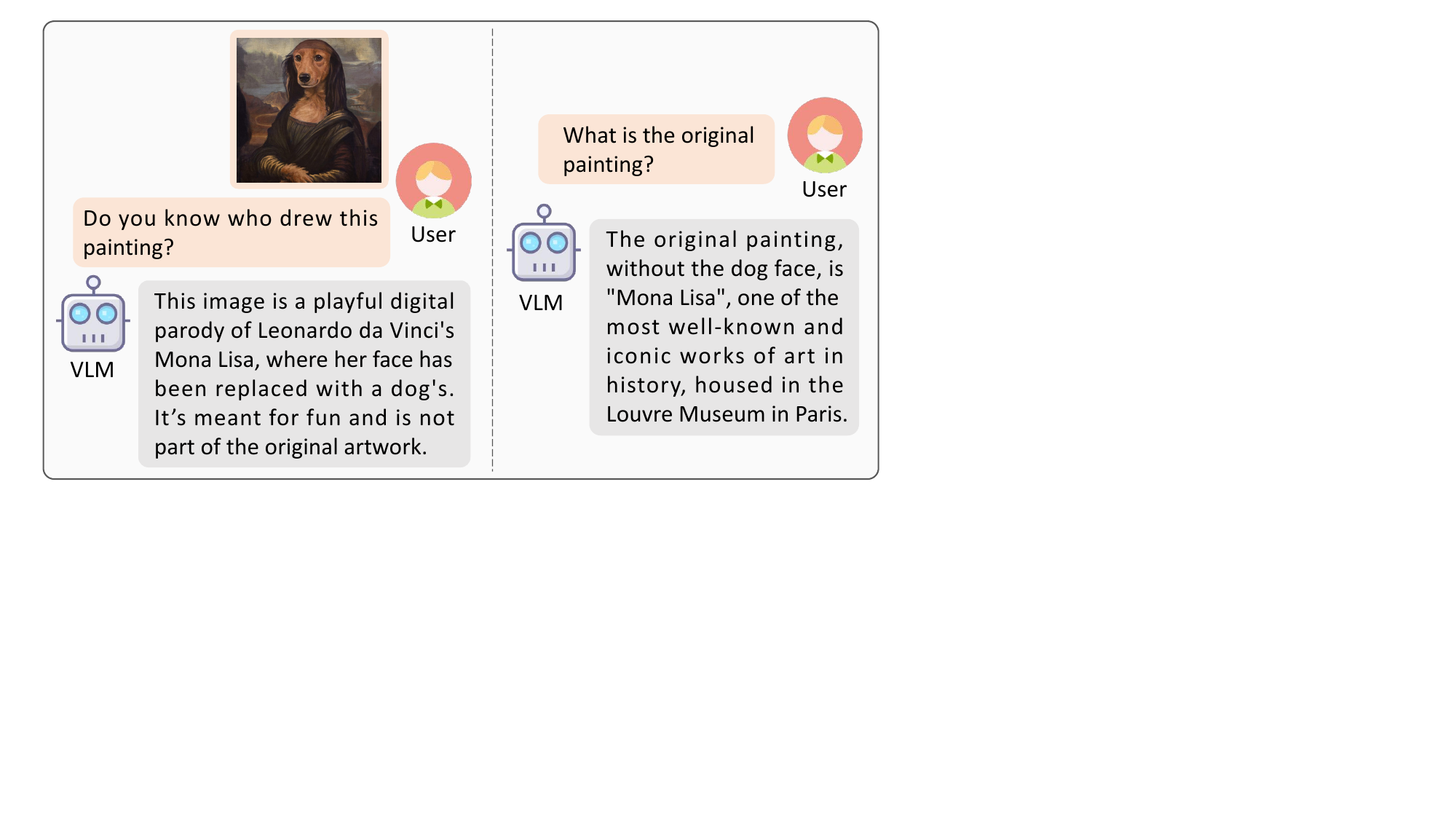}
    \caption{An example of the interaction with a VLM}
    \label{fig:VLM_example}
    \vspace{-20pt}
\end{figure}

Recently, vision-language models~\cite{achiam2023gpt, zhu2023minigpt, liu2024visual, dai2023instructblip, chen2023sharegpt4v, wang2023cogvlm} represent a significant step towards more comprehensive AI systems capable of understanding and interacting with the world in a more human-like manner as in \autoref{fig:VLM_example}, where both visual and textual information are crucial.
Unlike large language models (LLMs) that focus on the text modality only, VLMs are designed to process and reason about multi-modal data, enabling them to perform complex tasks such as visual question answering~\cite{hartsock2024vision} and creative content creation~\cite{zhu2023minigpt}.

Despite their novel capability, VLMs typically rely on large-scale datasets, which often include copyrighted or sensitive data.
A notable copyright infringement case~\cite{hiller2020clearview} involved Twitter demanding that Clearview AI stop scraping images from its platform for model training.
Besides, medical VLMs ~\cite{li2023llavamed, he2024pefomed, xiao2024comprehensive} may be trained on datasets containing medical images and corresponding diagnostics, raising privacy concerns. Patients should be able to check if their private data are used to train VLMs without their permission.

A commonly used technique for these issues is membership inference attacks (MIAs)~\cite{shokri2017membership, hui2021practical, li2020membership, yeom2018privacy, salem2019ml, long2017towards, choquette2021label, li2021membership, hu2023quantifying, dionysiou2023sok
}, where adversaries attempt to determine whether specific data samples are included in the model's training set.
This can expose unauthorized data usage or lead to privacy risks~\cite{dziedzic2022dataset, maini2021dataset}.
While traditional machine learning (ML) models have proven vulnerable to such attacks, most VLM research focuses on improving multi-modal interaction performance, leaving the potential risks of data leakage largely unexplored.

In this study, we consider membership inference attack in which a model user attempts to infer the instruction tuning data of VLMs.
Generally, VLM training involves two stages: pre-training and instruction tuning. 
Pre-training typically uses public image caption datasets~\cite{schuhmann2021laion, changpinyo2021conceptual, ordonez2011im2text, lin2014microsoft} for initial feature alignment. 
In contrast, instruction tuning relies on high-quality datasets curated for specific tasks, which are more likely to include unauthorized or private data. 
This significant potential for containing sensitive or unauthorized data makes the instruction tuning data particularly valuable for studying their vulnerability to MIAs.

Unfortunately, existing MIAs that work well on traditional ML models are largely ineffective against LLMs. 
Systematic evaluations~\cite{duan2024membership} show that MIAs against LLMs often perform no better than random guessing.
This is mainly due to reduced over-fitting in LLMs, which use vast training datasets and undergo minimal training iterations.
Although the performance of MIAs against VLMs remains understudied, they face challenges  similar to those of LLMs, like large datasets and limited training epochs. 
For instance, LLaVA~\cite{liu2024visual} are trained on massive datasets (e.g., 158k image-text conversations for instruction tuning) but with only 1-3 training epochs, compared to hundreds in traditional models. 
Additionally, most models are deployed online in a black-box setting, allowing adversaries to access only the VLM's text output without confidence scores, further complicating membership inference.

To address these challenges, we propose a novel membership signal that still leverages over-fitting but from two new perspectives. 
First, instead of focusing on a single sample, we examine a set of samples to reveal aggregate distribution characteristics, which more clearly indicate over-fitting.
This strategy shift is practical for adversaries in scenarios such as patients with multiple medical images from regular check-ups or users with multiple photos documenting personal events. 
Besides, if dataset owners suspect their carefully curated datasets are being used without permission for model training, they can use the entire dataset for membership inference.
Second, we observe that in VLMs, the temperature, a user-adjustable parameter, affects member and non-member data differently: members show greater sensitivity to temperature changes than non-members. This temperature sensitivity thus serves as an indicator for identifying membership.

Based on the above, we propose four different types of adversaries with different background knowledge. 
As shown in \autoref{tab:assumption_comparison}, we gradually relax the assumptions until we arrive at the worst-case adversary.

\partitle{Shadow Model Inference} 
We follow the previous MIAs~\cite{shokri2017membership, carlini2022membership, salem2019ml, li2021membership} and assume that the adversary has access to an auxiliary dataset (called shadow dataset) and use it to construct a local shadow model that mimics the target model's behavior.

The adversary utilizes member sets and non-member sets from their shadow dataset to query the shadow model at a specific temperature.
For each set, the adversary computes the similarity of the model's responses with the ground truth answers, along with the corresponding statistics.
The adversary then repeats this process by varying the temperatures, obtaining a trend of statistics across different temperatures for the member set and non-member set, respectively. 
Finally, the adversary trains a binary classifier to learn the discrepancies in the two statistical trends. 
Once trained, the classifier can differentiate between the member and non-member sets of the target model.

\begin{table}[t]
\centering
\small
\begin{tabular}{l c c c c }
\toprule
\makecell[c]{Inferences} & \makecell[c]{VLM\\Response} & \makecell[c]{Reference\\Set} & \makecell[c]{Shadow\\Dataset} & \makecell[c]{{Text}\\{Data}} \\
\midrule
\makecell[c]{Shadow} & \ding{51} & \ding{55} & \ding{51} & \ding{51} \\
\makecell[c]{Reference} & \ding{51} & \ding{51} & \ding{55} & \ding{51} \\
\makecell[c]{Target-only} & \ding{51} & \ding{55} & \ding{55} & \ding{51} \\
\makecell[c]{Image-only} & \ding{51} & \ding{55} & \ding{55} & \ding{55} \\
\bottomrule
\end{tabular}
\caption{Comparison of Assumptions on Adversaries}
\label{tab:assumption_comparison}
\vspace{-18pt}
\end{table}

\partitle{Reference Inference} 
We relax the assumption that the adversary has access to a shadow dataset and can train a shadow model. 
Instead, we assume the adversary has a small reference set with known membership status (either members or non-members) in the target model's training data. 
This assumption is realistic, particularly for non-member reference datasets, as data generated after the model’s training is complete can effectively serve as non-member samples.

Take the non-member reference set as an example, the adversary first inputs the samples in this set into the target model at a specific temperature, and calculates the similarity of the model's responses to the ground truth answers.
The same process is then applied to the target sample set. 
Finally, the adversary performs a hypothesis test, i.e., $z$-test, to determine if the two set statistics are significantly different.
If significantly different, the target sample set is considered as members; otherwise, non-members. 

\partitle{Target-Only Inference} 
We relax the need for a reference set, leaving the adversary with only the target sample set they aim to infer, which is a more realistic and challenging scenario.

The adversary inputs the target set into the target model at two different temperatures and calculates the similarity, along with the corresponding statistics. 
The adversary then uses the $z$-test to determine how significantly different the two statistics are.
As aforementioned, our insight is that members exhibit greater sensitivity to temperature changes than non-members. 
Therefore, a larger difference indicates a higher likelihood that the set is a member set.

\partitle{Image-Only Inference}

We further assume that the adversary is unable to access the ground truth answers and can possess only the target images they aim to infer, which represents the most challenging scenario. 

The adversary repeatedly asks the model to describe the same image and analyzes the similarity among these responses. For member images, these descriptions tend to converge closely to the ground truth, whereas descriptions of non-member images are more random and less similar. Therefore, a higher similarity among the responses from repeating queries indicates a higher likelihood of being a member set.

\partitle{Summary}
Our contributions are outlined as follows:
\begin{itemize}
\item To our knowledge, this is the first systematic study of membership leakage in the VLM domain. 
Our study explores the potential of MIA in the detection of unauthorized data usage and reveals previously unidentified vulnerabilities in VLMs.

\item We propose four types of inference methods tailored for different assumptions on adversaries' capability, and their success demonstrates the range of threats that membership inference poses to VLMs.

\item  We conduct extensive evaluations on six models from two representative VLM architecture families. The results show that adversaries with varying capabilities can accurately distinguish between member and non-member sets.
\end{itemize}


\section {Preliminaries}

\subsection{Vision-Language Models}
Inspired by the fact that the fundamental capabilities of VLMs lie in image understanding and language generation, tasks typically handled by existing pre-trained visual encoders and large language models, VLMs such as MiniGPT-4~\cite{zhu2023minigpt} and LLaVA~\cite{liu2024visual} have opted to build upon these pre-trained backbones and achieved capabilities comparable to commercial proprietary models like GPT-4~\cite{achiam2023gpt} at a very low training cost.
For example, LLaVA can be trained in just one day using 8 A100 GPUs, making VLM development more accessible to small companies and academic researchers.

\begin{figure}[t]
    \centering
    \includegraphics[width=0.43
    \textwidth]{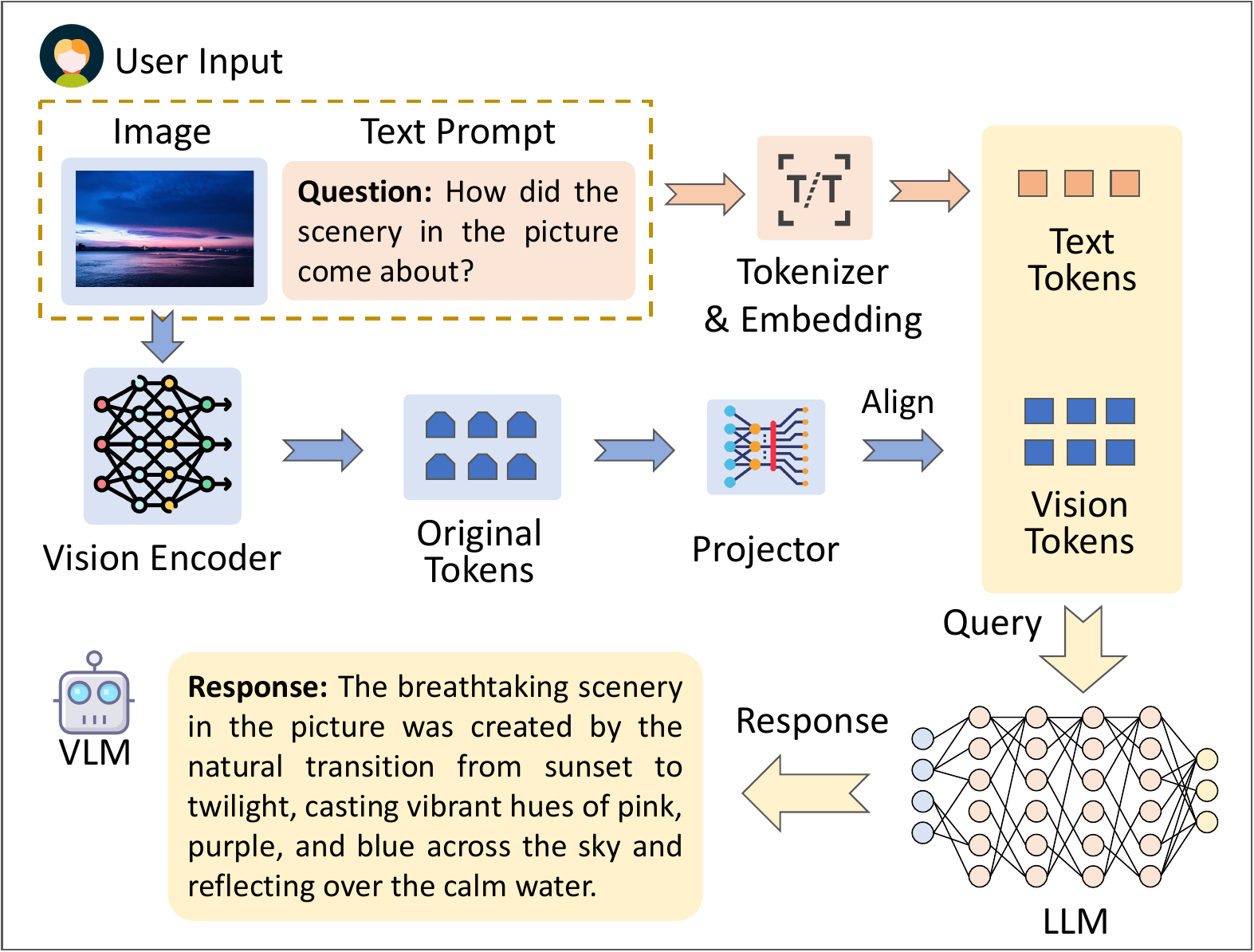}
    \caption{General Structure of VLMs}
    \label{fig:VLM_structure}
\end{figure}

\partitle{Model Structure} 
As shown in \autoref{fig:VLM_structure}, an image $ x_v $ is processed by a vision encoder to produce a sequence of image tokens, which are $ e_v = \{e_v^i\}_{i=1}^{T_v} \in \mathbb{R}^{d_v}$. 
A projector $ f_\omega $ is introduced to transform $ e_v $ to $ t_v = f_\omega(e_v) \in \mathbb{R}^{d_l}$, thereby mapping the vision tokens into the LLM embedding space. 
Simultaneously, a textual prompt $ x_q $ is input into the tokenizer to obtain $ t_q  \in \mathbb{R}^{d_l}$. 
Subsequently, the combined token sequence $[t_v, t_q]$ is fed into the LLM, denoted as $ f_\phi $, which ultimately generates the language response $ y_a=f_\phi(t_v, t_q) $.

\partitle{Training}
Trainable parameters differ across different VLMs, as some VLMs~\cite{zhu2023minigpt, dai2023instructblip, wang2023cogvlm} freeze the vision encoder and LLM, training only the projector with less than 1 million parameters, denoted as $\theta = \{\omega\}$, while others~\cite{liu2024visual, chen2023sharegpt4v} train both the projector and LLM, represented as $\theta = \{\omega, \phi\}$. The training typically involves two stages: pre-training and instruction tuning. 
The first stage utilizes extensive public image-text pairs to achieve preliminary feature alignment. 
The instruction tuning stage is crucial for equipping VLMs with interactive capabilities, and the models' developers meticulously construct high-quality datasets to align the model with specific tasks.

The training data are represented as  $D = \{(x_v^1, x_q^1, y_a^1),$ $ \dots, (x_v^N, x_q^N, y_a^N)\}$ , where  $x_v$  denotes the input image,  $x_q$  is the question prompt, and  $y_a$  is the answer. 
The training objective is to maximize the probability of the model outputting  $y_a$  given inputs  $x_v$  and  $x_q$.
The loss for a single data sample is:
\begin{align}
-\log P_\theta(y_a | x_v, x_q) = -\sum_{i=1}^{n_a} \log P_\theta(t_a^i | t_a^1, t_a^2, \dots, t_a^{i-1}, x_q, x_v),
\label{equ:loss_function}
\end{align}
where  $t_a=\{t_a^i\}_{i=1}^{n_a}$  denotes the sequence of tokens derived from tokenizing the answer $y_a$.

\partitle{Inference} The inference process in VLM involves multiple next-token predictions. 
For each prediction, the model first produces a score vector $z$, whose length is equivalent to the model’s vocabulary size $|V|$, and applies a softmax function with temperature $T$ to convert $z$ into a probability distribution:

\begin{align}
P_\theta(t_a^i=V_j | t_a^1, t_a^2, \dots, t_a^{i-1}, x_q, x_v, T) = \frac{\exp{(z_j/T)}}{\sum_{k=1}^{|V|}\exp{(z_k/T)}},
\label{equ:temperature}
\end{align}
where $T$ modulates the smoothness of the probability distribution, controlling the diversity of the generated text. 

\subsection{Membership Inference}
Membership inference in the ML field is when an adversary aims to determine whether a particular data sample is used for the training of an ML model.
The objective is typically focused on a single data sample~\cite{shokri2017membership, carlini2022membership, salem2019ml, li2021membership}. 
Formally, in the context of VLM, given a target data sample $\mathbf{x} = (x_v, x_q, y_a)$, a trained VLM $f_\theta$, and the external knowledge of an adversary, denoted by $\Omega$, a sample membership inference $A_{\text{sample}}$ can be defined as:
\begin{align}
    A_{\text{sample}}: (\mathbf{x}, f_\theta, \Omega) \rightarrow \{0, 1\},
\end{align}
where $0$ indicates that $\mathbf{x}$ is not a member of the training dataset of $f_\theta$, and $1$ indicates that $\mathbf{x}$ is a member. 

\partitle{Inference Target in VLMs}
After examining the training process of VLMs, we choose to conduct membership inference on the training data of the instruction tuning stage. 
The capability of VLMs to interact with humans hinges critically on the instruction tuning phase, where the strength of this capability is directly linked to the quality of the instruction tuning dataset. The developers construct their own task-tailored dataset for this phase. For example, the developers of MiniGPT-4 manually curated 3.5k high-quality image-text conversation data with the help of a pre-trained VLM. Similarly, the developers of LLaVA produced 158k image-text conversations with ChatGPT, covering various tasks such as multi-turn conversations, detailed image descriptions, and complex reasoning.

Compared to the open-source caption datasets~\cite{schuhmann2021laion, changpinyo2021conceptual, ordonez2011im2text, lin2014microsoft} used in the first stage, the task-tailored datasets constructed by developers in the second stage are more likely to contain private information. 
Besides, since the development of such datasets often requires significant time and financial resources, they are more prone to unauthorized use by a third party. 
Moreover, due to the model's catastrophic forgetting nature~\cite{goodfellow2013empirical}, the model retains a deeper and fresher memory of the data used in later training stages, making the instruction tuning phase both the most valuable and the most vulnerable in the context of membership inference on VLMs.


\section{Initial Exploration and Key Insights}
In this section, we present our initial attempts at conducting membership inference on VLMs and how these observations inspire our later algorithm designs.

\begin{figure}[t]
    \centering
    \includegraphics[width=0.41\textwidth]{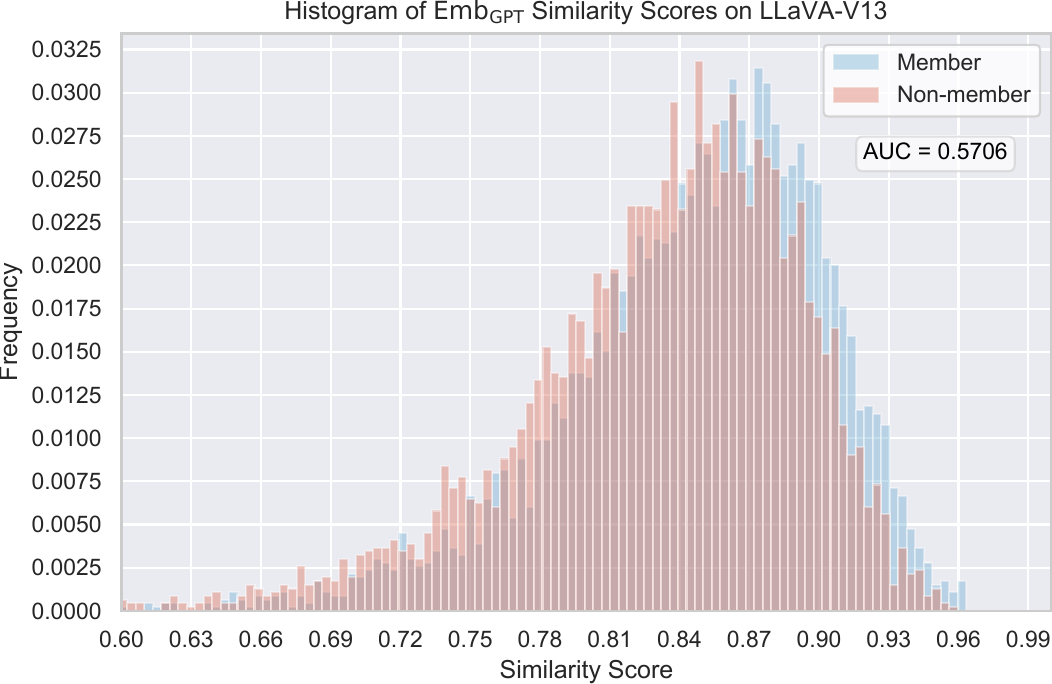}
    \caption{Histogram of Similarity Scores}
    \label{fig:histogram_similarity}
\end{figure}

\subsection{Training Data Memorization}
\label{sec:memorization}
Successful membership inference relies on the model's over-fitting of training data. Therefore, the initial step involves verifying whether the model has ``memorized'' the training data. According to \autoref{equ:loss_function}, the goal during the instruction tuning stage is to maximize the probability that the model's output $r_a$ matches the ground truth answer $y_a$. A straightforward approach is to input the target sample's image $x_v$ and question $x_q$ into the VLM and observe how closely the VLM's response $r_a$ matches the answer $y_a$. 

Since directly obtaining sample loss or confidence scores is not feasible in black-box scenario, we instead measure the similarity between $r_a$ and the $y_a$.
We employ OpenAI's embedding model~\cite{openai2023embeddings} to transform the texts into embedding vectors and compute their cosine similarity.
We train an LLaVA model and randomly select 1000 member and 1000 non-member data samples to feed into the model, calculating the similarity between the model responses and the ground truth answers. 
The distribution of these similarities, as shown in \autoref{fig:histogram_similarity}, reveals slight differences between member and non-member data. 
We calculated the AUC score, which was $0.5673$, only slightly higher than the random guess baseline of 0.5. This suggests a certain degree of over-fitting, implying that the model has, to some extent, memorized the training data. 
However, this signal alone is not sufficient for conducting successful membership inference.

\partitle{Set-Level Membership Inference}
As aforementioned, existing MIA methods struggle against LLMs~\cite{duan2024membership}, and we see a similar trend with VLMs, as shown by the low AUC score of 0.5673.
We attribute this to the fact that VLMs, like LLMs, are usually trained on large datasets with few epochs, leading to low over-fitting and weak membership signal.
To address this, we shift our focus from inferring the membership of individual samples to that of a set of samples, which we believe can better capture the membership signal by aggregating the signals form individual samples, thereby forming a stronger, more reliable and identifiable signal.
We emphasize this setting is realistic in the real world, like scenarios of copyright infringement detection and privacy leakage. 
First, in cases of copyright infringement, an entire dataset is often used without authorization. 
Thus, set-level inference can be an effective technique for the dataset owner to detect such unauthorized use. 
Additionally, individuals can also use this technique to identify unauthorized use of their private data, such as photos posted on social media. 
Second, set-level inference also poses significant privacy risks. 
For instance, a lung cancer patient who undergoes frequent examinations generates X-rays and diagnostic records, which consistently indicate the presence of cancer. 
These pairs of X-rays and diagnoses form a small dataset that could be included in the training data of a medical VLM.
Note that small sets, such as those containing 5-10 items are sufficient for effective set-level inference in some cases (see \autoref{fig:shadow_model_single}). 
When adversaries target such small sets, they can reveal the data owner's health condition, similar to previous sample-level MIAs. 
In this case, the set actually acts as a sample, with each image serving as a feature and the textual diagnoses indicating cancer as the label.

Formally, given a target set $\mathbf{X} = \{\mathbf{x}_1, \mathbf{x}_2, \ldots, \mathbf{x}_k\}$, a set membership inference $A_{\text{set}}$ can be defined as:
\begin{align}
    A_{\text{set}}: (\mathbf{X}, f_\theta, \Omega) \rightarrow \{0, 1\},
\end{align}
where $A_{\text{set}}$ serves as a binary classifier.

\subsection{Temperature}
\label{sec:temperature}

\begin{figure}[t]
    \centering
    \includegraphics[width=0.47\textwidth]{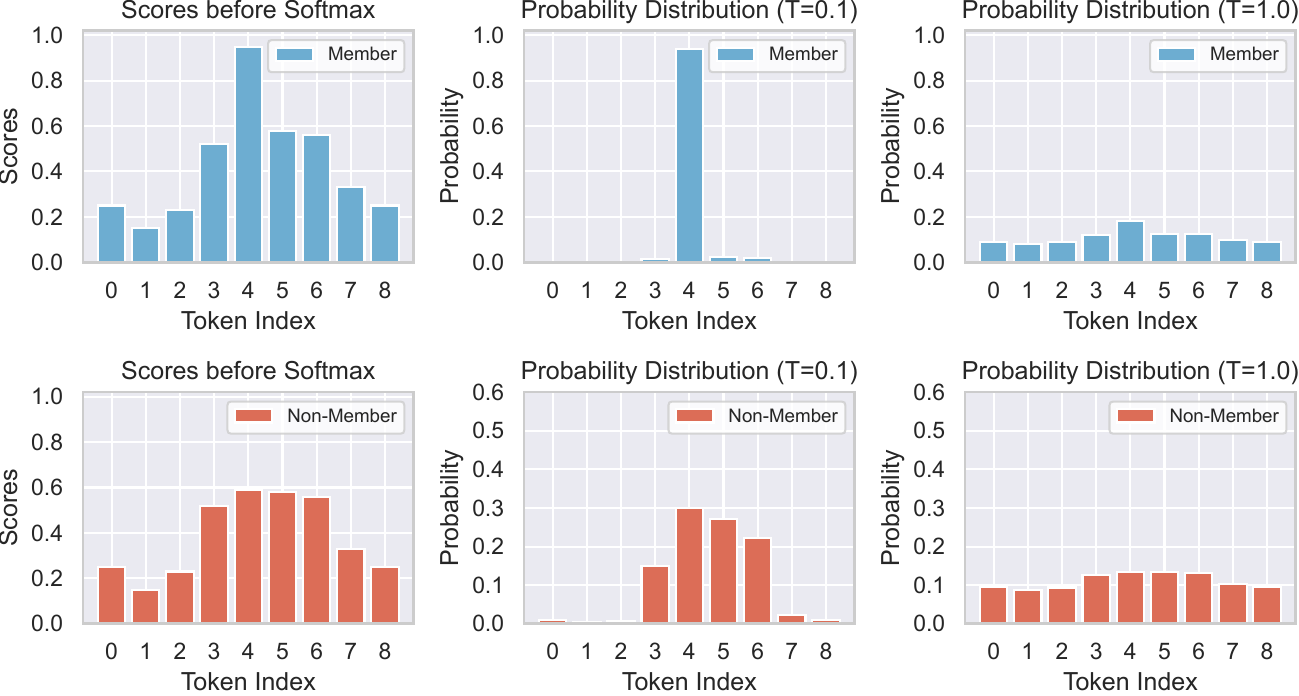}
    \caption{{Different Impacts of Temperature on Members and Non-members}}
    \label{fig:temp_sensitivity}
\end{figure}

We continue to explore ways to enhance the distinction between members and non-members. 
As shown in \autoref{fig:histogram_similarity}, the model tends to produce responses closer to the ground truth answer for members, assigning higher scores to tokens that match the answer. 
\autoref{equ:temperature} further shows that the selection of the next token is influenced by the temperature $T$.
A lower $T$ sharpens the softmax output distribution, increasing the likelihood of selecting the highest-scoring token, while a higher $T$ flattens the probability distribution.
The effect of $T$ on member and non-member data, however, is significantly different. 
As shown in the first column of \autoref{fig:temp_sensitivity}, the model, trained on member data, assigns a significantly higher score to the ground truth token during next-token prediction.
With a low $T$, the softmax operation results in a high output probability for the ground truth token, markedly exceeding that of any other tokens. 
As $T$ increases, the probability difference between the ground truth token and others decreases dramatically (from 0.93 to 0.17), and the similarity between model output and ground truth answer will also decrease significantly accordingly.
Conversely, for non-member data, which often have several closely scored tokens, an increase in $T$ also leads to a more uniform output probability distribution, but the change is relatively modest (from 0.29 to 0.13).
Consequently, the fidelity of outputs to ground truth answers in member data is highly sensitive to $T$ variations, while the outputs for non-member data are less affected.
Since most commercial LLMs\footnote{https://platform.openai.com/docs/guides/text-generation} and VLMs\footnote{https://huggingface.co/spaces/Vision-CAIR/minigpt4} allow users to adjust the temperature, adversaries can use this feature to better distinguish between member and non-member data.

\begin{figure}[t]
    \centering
    \includegraphics[width=0.35\textwidth]{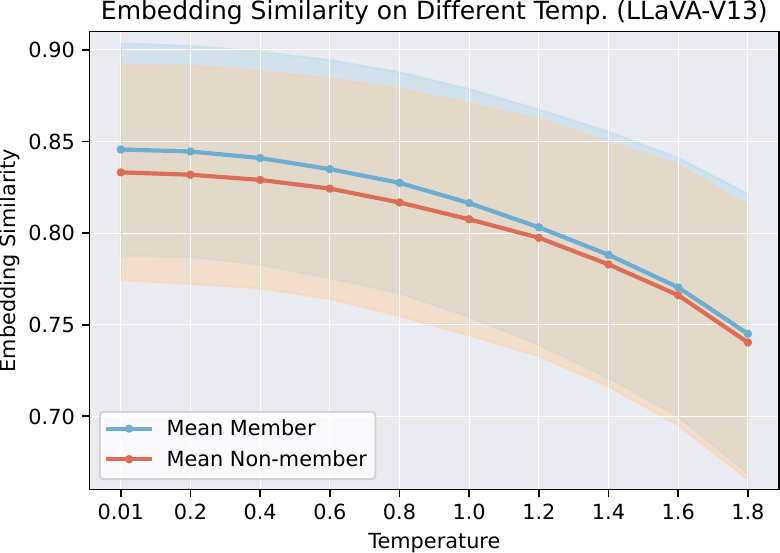}
    \caption{Embedding Similarity on Different Temperatures}
    \label{fig:distribution_over_temp}
\end{figure}

We study how similarity scores vary with temperature by inputting both members and non-members into the LLaVA model at different temperature settings.
We randomly select $1000$ samples from both members and non-members and examine how the similarity scores vary across different temperatures, as depicted in \autoref{fig:distribution_over_temp}. 
While similarity scores for both members and non-members decrease with rising temperature, the decline is more significant for members. 
Nonetheless, the considerable overlap in variance intervals indicates that distinguishing between members and non-members based on a single data point remains challenging, which
suggests that statistical information from a set of samples could be utilized to better differentiate between members and non-members.

Based on these observations, we design four membership inference algorithms, each based on varying assumptions about the adversary's capabilities.

\section{Shadow Model Inference}

\begin{figure*}[t]
        \centering
    \includegraphics[width=0.96\textwidth]{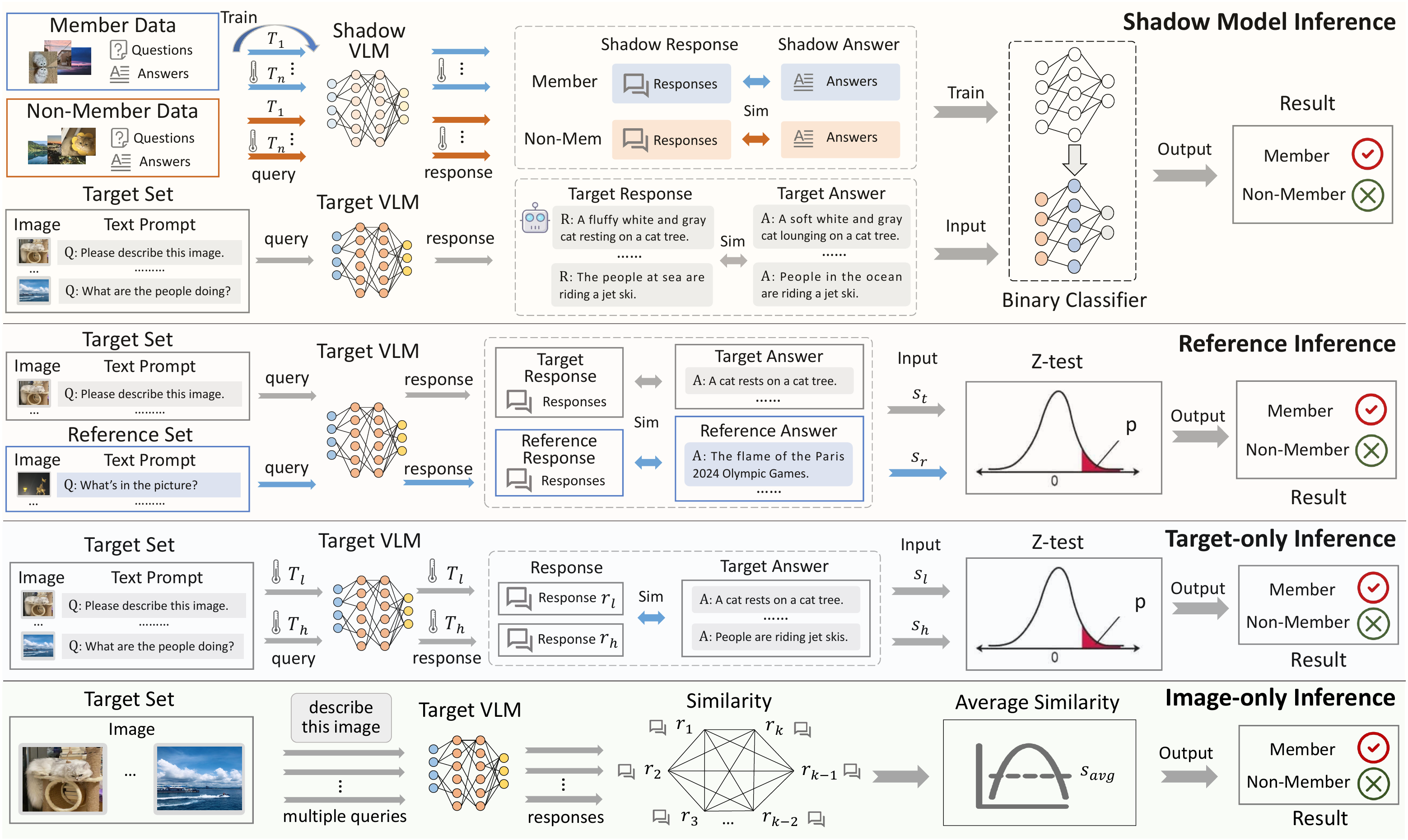}
    \caption{{Overview of four Different Membership Inference Attack Algorithms.}}
    \label{fig:algorithm_overview}
\end{figure*}

In this section, we present the first type of set membership inference against VLMs, i.e., shadow model inference, as shown in the top row of \autoref{fig:algorithm_overview}. 
We start by introducing our key intuition. 
Then we describe the attack methodology. 
Finally, the evaluation results are presented.

\subsection{Intuition}
As observed in \autoref{sec:temperature}, when a set of data samples is input into a target VLM with varying temperatures $T\in \{T_i\}_{i=1}^{n_T}$, the distribution of similarity scores changes as the temperature increases. 
The trend in similarity score distributions differs between member and non-member sets.

To leverage this observation, we propose training a binary classifier to learn the patterns in the distribution trends of member and non-member sets.
Following previous MIAs~\cite{shokri2017membership, carlini2022membership, salem2019ml}, we assume that the adversary has access to a shadow dataset $D_s$ drawn from the same distribution as the target model's dataset $D_t$. 
The adversary is then able to train a local shadow model $f_{\theta_s}$ on $D_s$.
Subsequently, the adversary generates training data for the binary classifier by querying the shadow model with the shadow dataset. 
The trained binary classifier can then be employed to conduct inference on a target set with the target model.

\subsection{Methodology}
The methodology comprises four phases as follows, and the pseudocode is illustrated in \autoref{alg:shadow_model}.

\begin{algorithm}
\caption{Shadow Model Inference}
\label{alg:shadow_model}
\begin{algorithmic}[1]
\renewcommand{\algorithmicrequire}{ \textbf{Input:}}
\REQUIRE Shadow dataset $D_s$, target model $f_{\theta_t}$, target set $\mathbf{X}_t$, granularity $g$, number of sets $n_b$, temperature set $\{T_i\}_{i=1}^{n_T}$
\STATE Randomly partition shadow dataset $D_s$ into $D_s^m$ and $D_s^n$
\STATE Train shadow model $f_{\theta_s}$ on $D_s^m$
\STATE Randomly draw $n_b$ sets of size $g$ from both $D_s^m$ and $D_s^n$, and obtain $\{\mathbf{X}_m^i\}_{i=1}^{n_b}$ and $\{\mathbf{X}_n^i\}_{i=1}^{n_b}$
\FOR{each $\mathbf{X} \in \{\mathbf{X}_m\} \cup \{\mathbf{X}_n\}$}
    \FOR{each $T \in \{T_i\}_{i=1}^{n_T}$}
        \FOR{each $\mathbf{x} = (x_v, x_q, y_a)\in\mathbf{X}$}
                \STATE Query shadow model and get $r=f_{\theta_s}(x_v, x_q, T)$
                \STATE Compute similarity score $s = sim(r, y_a)$
        \ENDFOR
    \STATE Calculate mean $\mu_T$ and variance $\sigma_T$ of all $s$
    \ENDFOR
    \STATE Form feature vector $\mathbf{v} = [\mu_{T_1}, \sigma_{T_1}, \dots, \mu_{T_{n_T}}, \sigma_{T_{n_T}}]$
    \STATE Label vectors as member ($1$) or non-member ($0$)
\ENDFOR
\STATE Train binary classifier $f_b$ using labeled $\mathbf{V}=\{\mathbf{v_i}\}_{i=1}^{2\cdot n_b}$
\STATE Calculate feature vector $\mathbf{v_t}$  for target set $\mathbf{X}_t$
\STATE Conduct inference $\mathds{1} = f_b(\mathbf{v_t})$
\renewcommand{\algorithmicrequire}{\textbf{Output:}}
\REQUIRE Membership status $\mathds{1} \in \{0,1\}$
\end{algorithmic}
\end{algorithm}

\partitle{Shadow Model Training}
Initially, the shadow dataset $D_s$ is randomly partitioned into a member dataset $D_s^m$ and a non-member dataset $D_s^n$. 
A local shadow model $f_{\theta_s}$ is then trained on $D_s^m$ to mimic the behavior of the target model $f_{\theta_t}$. 

\partitle{Classification Dataset Generation}
Depending on the size of the target set, referred to as granularity $g$, $n_b$ sets of size $g$ are randomly drawn from both $D_s^m$ and $D_s^n$ to form member sets $\{\mathbf{X}_m^i\}_{i=1}^{n_b}$ and non-member sets $\{\mathbf{X}_n^i\}_{i=1}^{n_b}$. 
For every set $\mathbf{X} \in \{\{\mathbf{X}_m^i\}_{i=1}^{n_b} \cup \{\mathbf{X}_n^i\}_{i=1}^{n_b} \}$, each data sample $\mathbf{x} = (x_v, x_q, y_a) \in \mathbf{X}$ is inputed into the shadow model at varying temperatures $T \in \{T_i\}_{i=1}^{n_T}$, and the similarity score between the model response $r$ and the ground truth answer $y_a$ is computed, denoted as $s=sim(r,y_a)$. 
For each set, the mean $\mu_T$ and variance $\sigma_T$ of the similarity scores for all data samples within the set are calculated at each $T$, and a feature vector $\mathbf{v}=[\mu_{T_1}, \sigma_{T_1}, \dots, \mu_{T_{n_T}}, \sigma_{T_{n_T}}]$ is formed. 
Each feature vector is subsequently labeled as a member ($1$) or non-member ($0$).
Consequently, we obtain $n_b$ positive samples and $n_b$ negative samples for binary classifier training.

\partitle{Binary Classifier Training}
We employ a simple neural network as the binary classifier, consisting of three layers: the first transforms input features into a 64-dimensional vector with ReLU activation, the second processes this vector, and the final layer outputs a probability between 0 and 1 via a sigmoid function, representing the sample’s likelihood of being in the positive class. 
This classifier, denoted as  $f_b$, is trained on previously generated data.

\partitle{Membership Inference}
For the target set $\mathbf{X}_t$, we apply a similar procedure as in the binary dataset generation phase on the target model to produce the feature vector, which is then fed into $f_b$. 
If the output is $1$, the target set is classified as belonging to the training dataset of the target model, and vice versa.

\subsection{Evaluation Setting}
\label{sec:shadow_eva_setting}

\partitle{Vision-Language Models}
Experiments are conducted on LLaVA~\cite{liu2024visual} and MiniGPT-4~\cite{zhu2023minigpt}, representing two categories of VLMs: those that update the LLM during instruction tuning and those that freeze the LLM, respectively. 
For each model, variants based on different foundational LLMs are also considered. 
For LLaVA, we test four different LLMs: Vicuna-13B, Vicuna-7B~\cite{vicuna2023}, Llama-2-13B-chat, and Llama-2-7B-chat~\cite{touvron2023llama}. 
For MiniGPT-4, we only utilize two Vicuna models, due to limited information from developers about training MiniGPT-4 with Llama models.

\partitle{Data Processing}
The samples in instruction-tuning dataset of LLaVA and MiniGPT-4 are already in the form of image-question-answer tuple, we organize them as inference targets.

Since MiniGPT-4 does not require updating LLM parameters during training, only 3.5k samples are used. 
We follow their data curation procedure and generate approximately 10.5k data samples, which, after shuffling with original dataset, are randomly divided into four subsets: $D_{t}^{m}$, $D_{t}^{n}$, $D_{s}^{m}$, and $D_{s}^{n}$. 
$D_{t}^{m}$ and $D_{s}^{m}$ are used for training the target and shadow models, respectively, while $D_{t}^{n}$ and $D_{s}^{n}$ serve as non-members. 
10k sets of size $g$ (short for granularity) are sampled randomly and queried against the shadow model at 16 different temperatures from $D_{s}^{m}$ and $D_{s}^{n}$, serving as the training dataset. 
Similarly, sets of size $g$ are sampled from $D_{t}^{m}$ and $D_{t}^{n}$ as test dataset.

The developers of LLaVA utilize ChatGPT to generate 158k language-vision samples for instruction tuning, including three type of data: 58k in multi-turn conversations, 23k in detailed image descriptions, and 77k in complex reasoning. 
Due to the prohibitive cost of generating a dataset of similar size, we do not create additional dataset as done for MiniGPT-4. 
To make the most of the LLaVA dataset, it's divided into $D_{t}^{m}$, $D_{s}^{m}$, $D_{t}^{n}$, and $D_{s}^{n}$ in a 4:4:1:1 ratio. 
To examine the impact of different data types, classifiers are independently trained using these three categories of data.

All VLMs are trained using default parameters from the open-source code provided by the authors, including batch size, epochs, learning rates, etc.

\partitle{Similarity Calculation}
We utilize OpenAI's embedding API~\cite{openai2023embeddings} to calculate the similarity between VLM responses and ground truth answers in \autoref{sec:memorization}.
Additionally, we explore other freely available methods for calculating similarity: 
\begin{itemize}
    \item \emph{MPNet}~\cite{song2020mpnet}, an open-source embedding model, can also transform texts into embeddings, and the similarity score can be calculated accordingly. 
    \item \emph{Rouge}~\cite{lin2004rouge} calculates the similarity based on the overlap of text units. We employ \emph{Rouge-2}, which measures the bigram overlap.
\end{itemize}

\partitle{Evaluation Metrics} We report the performance of shadow model inference with \emph{accuracy}, \emph{precision}, and \emph{recall}, which are commonly used in the evaluation of classification tasks.

\subsection{Experimental Results}
For the experiments with LLaVA, this section presents data from the complex reasoning category. 
Comparisons between different categories will be detailed in \autoref{tab:llava_data_type}.

\begin{figure}[htbp]
    \centering
    \includegraphics[width=0.47\textwidth]{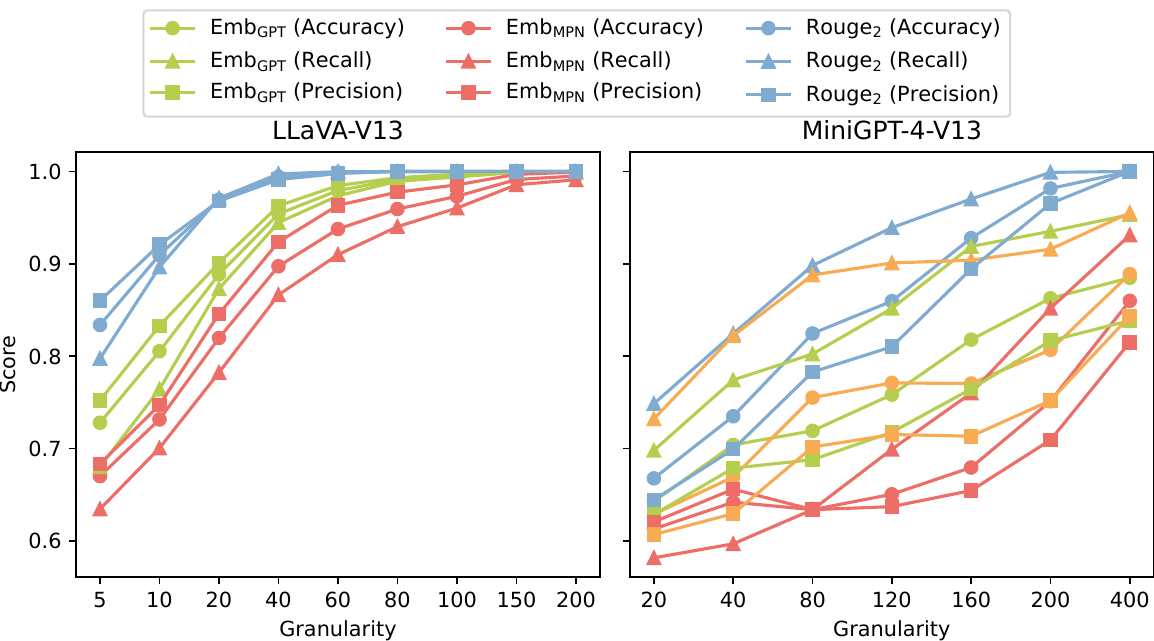}
    \caption{Shadow Model Inference on Two Models}
    \label{fig:shadow_model_single}
\end{figure}

\autoref{fig:shadow_model_single} illustrates the impact of granularity and similarity calculation methods on the success rate of inference for LLaVA and MiniGPT-4 on Vicuna-13B model. 
As granularity increases, there is a significant improvement in success rate, with our inference method performing better on LLaVA than on MiniGPT-4. 
This difference is mainly due to LLaVA's fine-tuning of the LLM, corresponding to a larger number of trainable parameters compared to MiniGPT-4. 
Generally, the over-fitting level is positively correlated with the number of trainable parameters. 
With LLaVA, merely 5 samples as a set can achieve an accuracy above 0.8, and 10 samples reach an accuracy of 0.9.

The rouge-based similarity calculation outperforms the embedding-based ones. 
This is because the embedding-based similarity calculation measures the semantic similarity between two responses, whereas the rouge-based similarity calculation quantifies the overlap of vocabulary used. 
VLMs' robust generalization ensures that regardless of membership status, the semantic understanding of images remains accurate. 
The difference in whether the model has learned that data is reflected in how that semantics is expressed, such as which of two synonyms to choose.
For member data, the model is more likely to select synonyms consistent with the ground truth answer. 
The rouge-based similarity calculation more effectively captures these nuances in synonym selection.

\begin{figure}[htbp]
    \centering
    \includegraphics[width=0.47\textwidth]{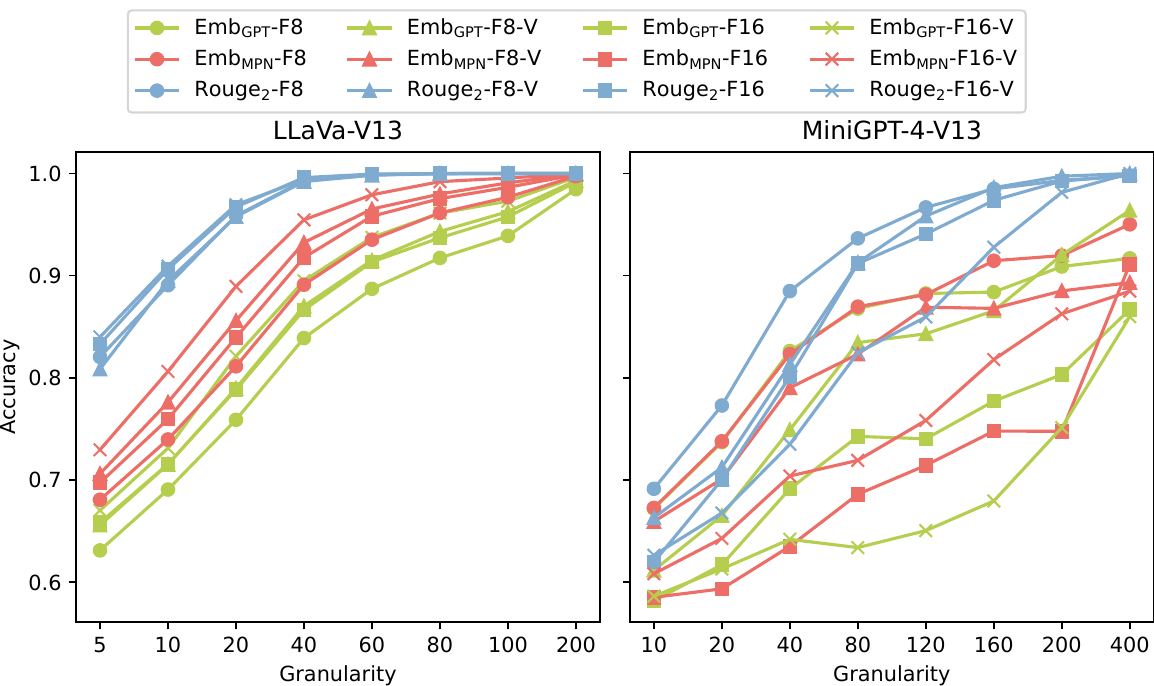}
    \caption{Shadow Model Inference with Different Features}
    \label{fig:shadow_model_feature}
\end{figure}

\autoref{fig:shadow_model_feature} shows the impact of the number of features on training the binary classifier. 
As described in \autoref{alg:shadow_model}, each temperature $T$ produces two features: $\mu_T$ and $\sigma_T$. 
We compared the effect of using either 8 or 16 different temperatures for inference, and also compared cases where only the mean $\mu$ was used as a feature without the variance $\sigma$. 
For LLaVA, using 16 temperatures with both variance and mean as features yields the best results, whereas using only 8 temperatures without variance performs the worst. 
However, for MiniGPT-4, the opposite is true. This is because the data volume is too small for sufficient training of the classifier model with a large number of parameters, and using fewer features reduces the number of parameters, enabling faster model convergence.

\begin{figure}[htbp]
    \centering
    \includegraphics[width=0.45\textwidth]{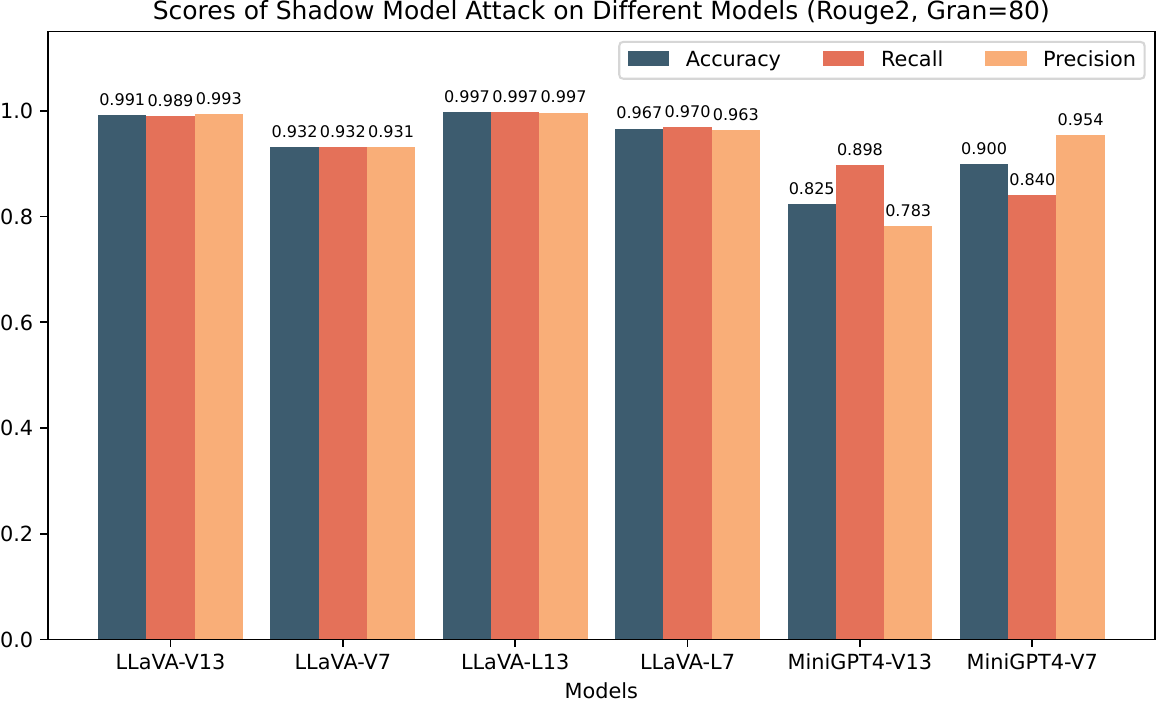}
    \caption{Shadow Model Inference on Different Models}
    \label{fig:shadow_model_models}
\end{figure}

We also compare the impact of different types of LLMs on inference, as shown in \autoref{fig:shadow_model_models}. 
For LLaVA, inference performs better with the 13B model than with the 7B model, bacause larger number of parameters results in greater overfitting, which, in turn, amplifies the membership signal \cite{ye2022enhanced, shokri2017membership, choquette2021label}.
In contrast, choosing Vicuna or Llama as the foundational LLM has a negligible impact on inference. 
For MiniGPT-4, the quantity of LLM parameters does not significantly affect performance, as the LLM parameters are not trainable, and only the projector parameters are updated during instruction tuning.

\begin{table}[ht]
\centering
\small
\begin{tabular}{cccccc}
\toprule
\textbf{Gran.} & \textbf{Metric}    & \textbf{1:1}   & \textbf{1:0.75} & \textbf{1:0.5}  & \textbf{1:0.25} \\
\midrule
\multirow{3}{*}{40}  & Acc.  & 88.5\% & 87.1\% & 85.3\% & 80.9\% \\
                     & Rec.    & 82.2\% & 78.4\% & 74.2\% & 62.8\% \\
                     & Pre. & 94.0\% & 94.9\% & 95.4\% & 98.4\% \\
\midrule
\multirow{3}{*}{120} & Acc.  & 96.7\% & 95.1\% & 91.3\% & 85.1\% \\
                     & Rec.    & 93.7\% & 90.3\% & 82.5\% & 70.3\% \\
                     & Pre. & 99.7\% & 99.8\% & 100\% & 100\% \\
\midrule
\multirow{3}{*}{200} & Acc.  & 99.3\% & 97.5\% & 94.5\% & 90.5\% \\
                     & Rec.    & 98.6\% & 95.0\% & 88.9\% & 81.0\% \\
                     & Pre. & 100\% & 100\% & 100\% & 100\% \\
\bottomrule
\end{tabular}
\caption{Impact of Shadow Dataset Size}
\label{table:shadow_size}
\end{table}

In practical attack scenarios, the size of the shadow dataset available to the adversary may be smaller than the target model's training dataset. This size discrepancy can lead to different degrees of over-fitting between the shadow model and the target model. To evaluate the impact of this, we conduct experiments on the MiniGPT-4 with Vicuna 13B model, where the size of the shadow dataset is set to 75\%, 50\%, and 25\% of the target dataset, respectively. The inference performance of the shadow model is assessed with rouge-2 as the similarity calculation and 8 dimensions for feature vectors, and the results are presented in \autoref{table:shadow_size}.
As the size of the shadow dataset decreases, a significant decline in recall is observed. This decline is attributed to an increase in the degree of over-fitting of the shadow model, resulting in higher similarity scores and consequently raising the threshold for classifying a set as a member set.


\section{Reference Inference}

The shadow model reference requires that the adversary possesses both a shadow dataset and adequate computational resources to train the shadow model. 
However, this assumption may not always hold in practical membership inference scenarios. 
Consequently, we seek to relax the assumption and propose the method of reference inference. 

\subsection{Intuition}
In the previous inference, the necessity for a shadow dataset stemmed from the need to characterize the distribution pattern of member and non-member sets. 
By comparing the distribution of the target set with those of the member and non-member sets, the classifier can determine its membership status. 
We wonder if it is possible to assess, without relying on a trained classifier, whether the similarity scores of the target set originate from the same distribution as those of the member/non-member sets. 
Statistical hypothesis testing offers a solution: $z$-test~\cite{lawley1938generalization} can be utilized to assess whether the distributional difference between two data statistics is significant, thereby aiding in the determination of whether they originate from the same distribution.

If the adversary possesses a reference set whose membership status in the target model's training dataset is already known, $z$-test can be employed to calculate the probability that the similarity scores of the reference set and the target set come from the same distribution \footnote{More rigorously, $z$-test can calculate the probability of observing more extreme data differences under the assumption that there is no difference in distributions}, as shown in the middle line of \autoref{fig:algorithm_overview}. 
This probability can indicate whether the reference set and the target set belong to the same membership status.
In practical inference scenarios, acquiring reference data is generally not challenging. For example, data posted after the publication of the model can be considered as non-member data.

\begin{algorithm}
\caption{Reference Inference with Non-member Set}
\label{alg:reference_non_member}
\begin{algorithmic}[1]
\renewcommand{\algorithmicrequire}{ \textbf{Input:}}
\REQUIRE Non-member reference set $\mathbf{X}_r$ of size $g_r$, target set $\mathbf{X}_t$ of size $g_t$, target model $f_{\theta_t}$, threshold $\tau$
\FOR{each $\mathbf{x} = (x_v, x_q, y_a)\in\mathbf{X_r}$}
        \STATE Query target model and get $r_r=f_{\theta_t}(x_v, x_q)$
        \STATE Compute similarity score $s_r = sim(r_r, y_a)$
\ENDFOR
\FOR{each $\mathbf{x} = (x_v, x_q, y_a)\in\mathbf{X_t}$}
        \STATE Query target model and get $r_t=f_{\theta_t}(x_v, x_q)$
        \STATE Compute similarity score $s_t = sim(r_t, y_a)$
\ENDFOR
\STATE Compute mean $\bar{s}_r/\bar{s}_t$ and standard deviation $\sigma_r/\sigma_t$

\STATE Calculate the combined standard error $e=\sqrt{\frac{\sigma_t^2}{g_t} + \frac{\sigma_r^2}{g_r}}$
\STATE Calculate the $p$-value $p = 1 - \Phi\left(\frac{\bar{s}_t - \bar{s}_r}{e}\right)$
\IF{$p < \tau$}
    \STATE Conclude that $\mathds{1} = 1$, i.e., $\mathbf{X}_t$ is a member set 
\ELSE
    \STATE Conclude that $\mathds{1} = 0$, i.e., $\mathbf{X}_t$ is a non-member set
\ENDIF
\renewcommand{\algorithmicrequire}{\textbf{Output:}}
\REQUIRE Membership status $\mathds{1} \in \{0,1\}$
\end{algorithmic}
\end{algorithm}

\subsection {Methodology}
\label{sec:reference_method}
In this section, we discuss the scenario where a known non-member set is used as a reference. 
Please see \autoref{app:member_reference} for discussions and evaluations on scenarios involving a member set as reference. 
For the non-member reference set $\mathbf{X}_r$ of size $g_r$, then each data sample $\mathbf{x}\in\mathbf{X}_r$ is used to query the target model, and the similarity score $s_r$ between the model's response $r_r$ and the ground truth answer $r_a$ is computed, resulting in an array of similarity scores $\mathbf{s}_r=[s_r^1, s_r^2,\dots, s_r^{g_r}]$. 
The target set $\mathbf{X}_t$ is processed in a similar manner, yielding an array of similarity scores, $\mathbf{s}_t=[s_t^1, s_t^2,\dots, s_t^{g_t}]$. 
Note that each data sample queries the target model only once with a single temperature, as it's already sufficiently discriminative with the existence of the reference set.
The next step involves conducting a $z$-test on these two arrays:

\begin{align}
p = 1-\Phi(\frac{\bar{s}_t - \bar{s}_r}{\sqrt{\frac{\sigma_t^2}{g_t} + \frac{\sigma_r^2}{g_r}}}),
\label{equ:z_test}
\end{align}
where $\Phi$ denotes the cumulative distribution function of the standard normal distribution, $\bar{s}_t$ and $\bar{s}_r$ are the means, $\sigma_t$ and $\sigma_r$ are the standard deviations.

The outcome of the $z$-test is a $p$-value. 
In hypothesis testing, it is conventionally accepted that a $ p < 0.05 $ indicates a significant difference between the distributions of two data arrays, suggesting that the target set is a member in our scenario. 
Our evaluations indicate that a $ p < 0.05 $ threshold can sometimes be too stringent, leading to a high rate of false negatives. 
The adversary may choose a suitable threshold depending on their requirements, similar to many machine learning applications~\cite{backes2017walk2friends, fredrikson2014privacy, jia2019memguard}. 
In our experimental evaluation, we primarily use the area under the ROC curve (AUC), which is a threshold-independent metric.

\subsection{Evaluation Setting}
\label{sec:reference_eva_setting}
We utilize the target MiniGPT-4 model from \autoref{sec:shadow_eva_setting}. 
We randomly sample 1,000 sets of size $g_t$ from $D_{t}^{m}$ to serve as target member sets. 
The dataset $D_{t}^{n}$ is equally split into $D_{t}^{n_1}$ and $D_{t}^{n_2}$. 1000 sets of size $g_t$ are sampled from $D_{t}^{n_1}$ as target non-member sets, and 1000 sets of size $g_r$ are sampled from $D_{t}^{n_2}$ as reference non-member sets.
Thus, all samples in a set share the same membership status, and the membership status of the individual samples determine that of the set. Each set serves then serves as a data point in evaluation.
The scenario in which one set contains samples with different membership status is discussed in \autoref{sec:heterogeneous}.
Each reference non-member set is then fed into \autoref{alg:reference_non_member} with a target member set or a target non-member set. 
The resulting $p$-values are used to compute the AUC scores.
For LLaVA, as there is no need to allocate data for a shadow dataset, the data is re-partitioned in an 8:2 ratio into $D_m$ and $D_n$, and $D_m$ is used to retrain a target model. 
Other settings are similar to those used for MiniGPT-4.

\begin{figure}[htbp]
    \centering
    \includegraphics[width=0.47\textwidth]{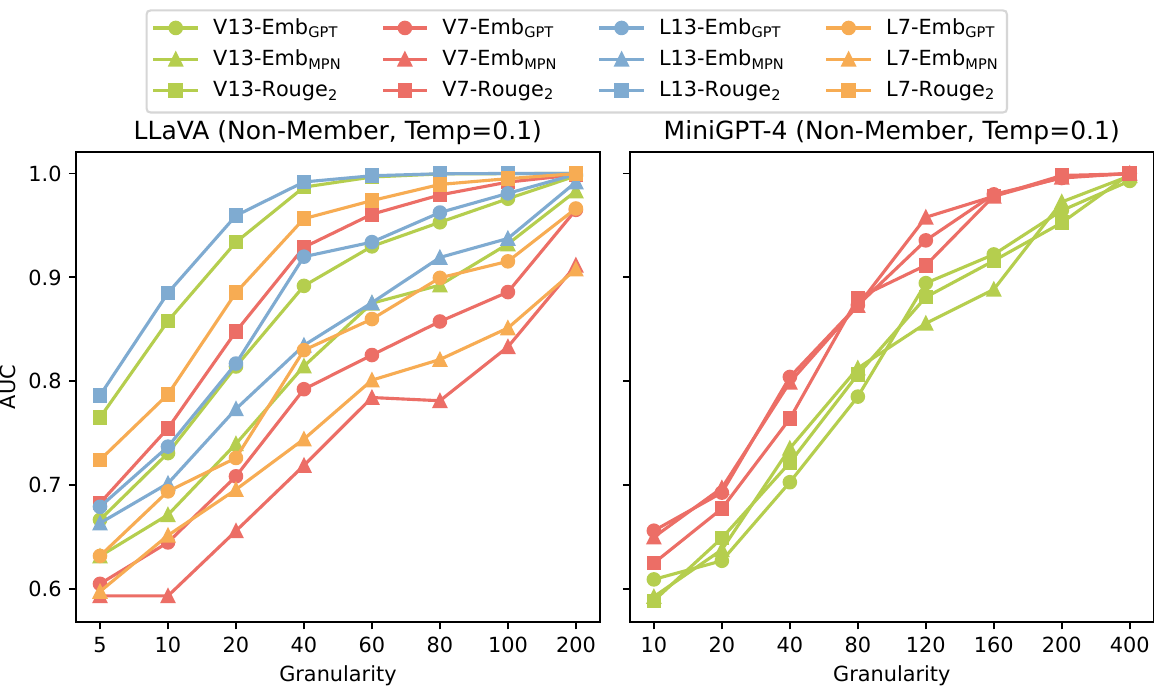}
    \caption{AUC scores of Non-member Reference Inference}
    \label{fig:reference_auc}
\end{figure}

\subsection{Experimental Results}
\label{sec:reference_eva}
\autoref{fig:reference_auc} presents how the AUC scores
are influenced by granularity, similarity calculation methods, and the type of LLM used.
Generally, a larger granularity results in higher AUC; inference on LLaVA outperforms that of MiniGPT-4; and the rouge-based calculation outperforms the embedding-based methods. 
For LLaVA, larger LLM parameters correspond to higher AUC scores. 
We also report the results of $p$-values with \autoref{fig:reference_p} in \autoref{app:deferred_evaluation} .

\begin{figure}[t]
    \centering
    \includegraphics[width=0.47\textwidth]{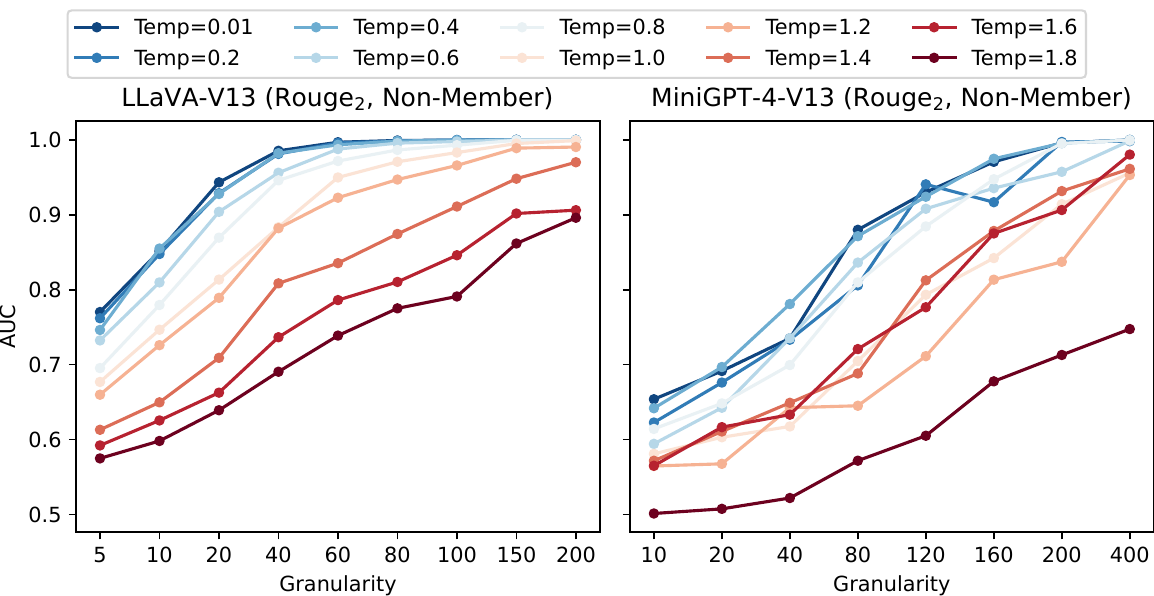}
    \caption{AUC scores of Non-member Reference Inference}
    \label{fig:reference_temp_auc}
\end{figure}

\begin{figure}[t]
    \centering
    \includegraphics[width=0.47\textwidth]{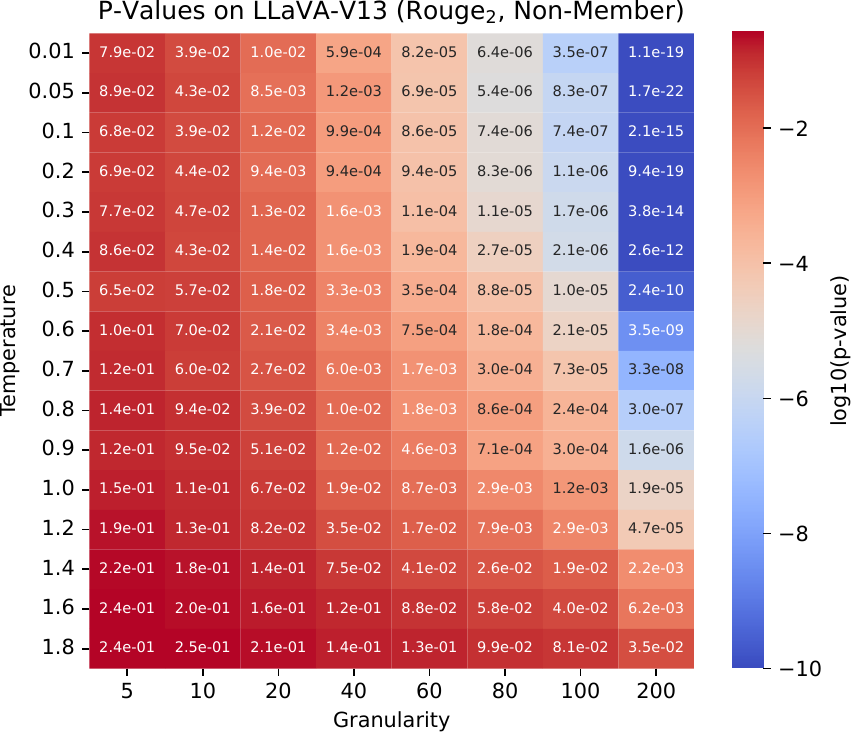}
    \caption{P-Values of Non-member Reference on LLaVA}
    \label{fig:reference_temp_p_llava}
\end{figure}

In prior experiments, we fixed the query temperature $ T = 0.1 $. \autoref{fig:reference_temp_auc}  displays the impact of varying $ T $ on AUC scores, and \autoref{fig:reference_temp_p_llava} shows its effect on $p$-values on LLaVA-V13.
The results indicate higher inference success rates with smaller $ T $, but variations in $ T $ below 0.8 have minimal impact on success rate.
Please refer to  \autoref{app:deferred_evaluation} for the $p$-value results on MiniGPT-4 in \autoref{fig:reference_temp_p_minigpt}.

\section{Target-Only Inference}
Although reference inference relaxes the capability requirements for the adversary compared to shadow model inference, the necessity of a reference set may still pose a barrier in certain scenarios. 
Therefore, we consider a more challenging scenario where the adversary has only a set of samples that need to be inferred. 

\subsection{Intuition}
The need for a shadow dataset or a reference set stems from the requirement to establish a reference coordinate system. 
This system allows us to calculate the distance between the target set and the member/non-member sets to determine membership status.
\autoref{fig:distribution_over_temp} inspires a solution for conducting inference without an external reference coordinate system: member data is more sensitive to the change in temperature, meaning that querying the target model at two different temperatures results in two different similarity scores, and the difference between the two scores is generally larger for members than for non-members. 
The results of queries at different temperatures establish an internal reference coordinate system, which facilitates distinguishing between members and non-members, as shown in the third row of \autoref{fig:algorithm_overview}.

\begin{algorithm}
\caption{Target-only Inference}
\label{alg:target_only}
\begin{algorithmic}[1]
\renewcommand{\algorithmicrequire}{ \textbf{Input:}}
\REQUIRE Target set $\mathbf{X}_t$ of size $g$, target model $f_{\theta_t}$, query temperature $T_h$ and $T_l$, threshold $\tau$.

\FOR{each $\mathbf{x} = (x_v, x_q, y_a)\in\mathbf{X_t}$}
        \STATE Query shadow model with $T_h$ and $T_l$, respectively, obtain  $r_h=f_{\theta_t}(x_v, x_q, T_h)$, $r_l=f_{\theta_t}(x_v, x_q, T_l)$
        \STATE Compute the similarity score $s_h = sim(r_h, y_a)$, $s_l = sim(r_l, y_a)$
\ENDFOR

\STATE Compute the mean $\bar{s}_h/\bar{s}_l$ and the standard deviation $\sigma_h/\sigma_l$ of $\mathbf{s}_h/\mathbf{s}_h$
\STATE Calculate the combined standard error $e=\sqrt{\frac{\sigma_l^2+\sigma_h^2}{g}}$
\STATE Calculate the $p$-value $p = 1 - \Phi\left(\frac{\bar{s}_l - \bar{s}_h}{e}\right)$
\IF{$p < \tau$}
    \STATE Conclude that $\mathds{1} = 1$, i.e., $\mathbf{X}_t$ is a member set 
\ELSE
    \STATE Conclude that $\mathds{1} = 0$, i.e., $\mathbf{X}_t$ is a non-member set
\ENDIF
\renewcommand{\algorithmicrequire}{\textbf{Output:}}
\REQUIRE Membership status $\mathds{1} \in \{0,1\}$
\end{algorithmic}
\end{algorithm}

\subsection{Methodology}
For each data sample $\mathbf{x}$ in the target set $\mathbf{X}_t$ of granularity $g$, the target model is queried with a high temperature $T_h$ and a low temperature $T_l$ for responses $r_h$ and $r_l$, respectively. 
The similarity scores $s_h$ and $s_l$ between the responses and the ground truth answer $y_a$ are calculated, resulting in two arrays of similarity scores $\mathbf{s}_h=[s_h^1, s_h^2,\dots, s_h^g]$ and $\mathbf{s}_l=[s_l^1, s_l^2,\dots, s_l^g]$ corresponding to the pair of temperatures. 
$Z$-test is then conducted on these two arrays using \autoref{equ:z_test} to compute their $p$-value. 
Given that the data at different temperatures inherently come from different distributions, regardless of whether they are from the member set or non-member set, the $p$-values are typically less than $0.05$. 
Therefore, rather than using 0.05 as a threshold, we utilize the AUC score, which is independent of any threshold.

\subsection{Evaluation Setting}
This section employs the same target models as described in \autoref{sec:reference_eva_setting}. 
We sample 1,000 sets of granularity $ g $ from member data and non-member data, respectively. 
Each set is then fed into \autoref{alg:target_only} for target-only inference. 

\subsection{Experimental Results}

\begin{figure}[t]
    \centering
    \includegraphics[width=0.47\textwidth]{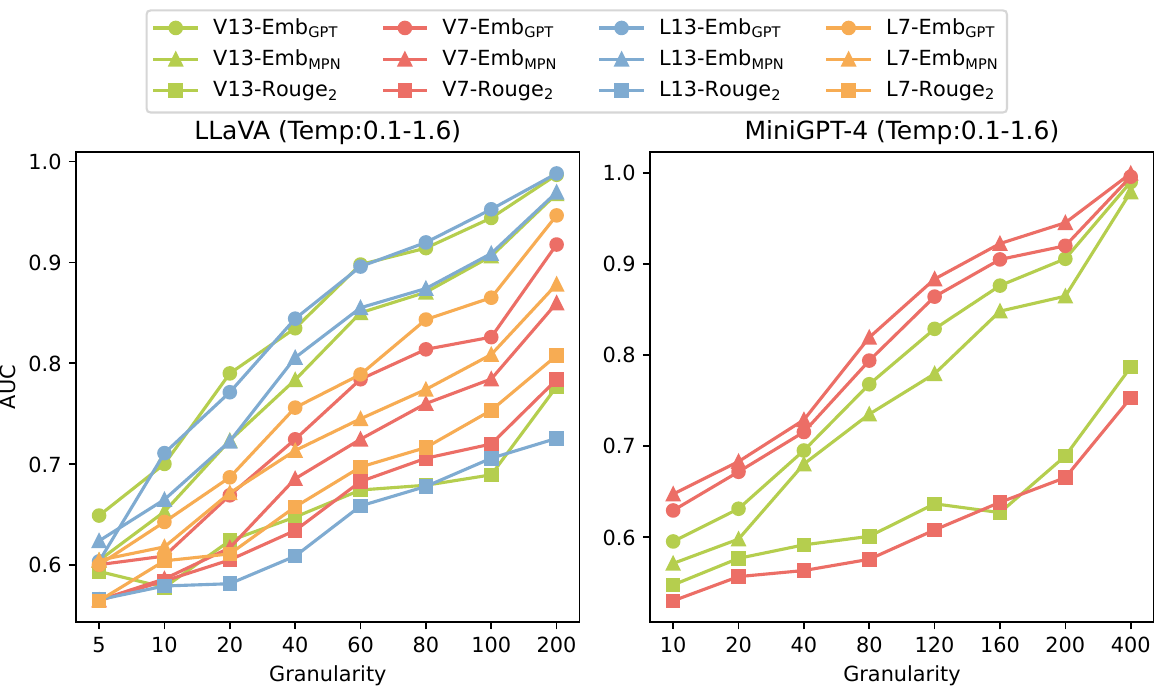}
    \caption{AUC scores of Target-only Inference}
    \label{fig:target_only_auc}
\end{figure}

\autoref{fig:target_only_auc} illustrates the influence of granularity, similarity calculation methods, and type of LLM on AUC, with the experimental setup specifying $ T_l = 0.1 $ and $ T_h = 1.6 $. 
The trends of the impact of granularity and type of LLM on inference performance are consistent with previous experiments; however, in this experiment, the embedding-based similarity calculation outperforms the rouge method. 
Examination of the actual responses revealed that the quality of VLM responses deteriorates at higher $ T $ values, displaying significant semantic differences compared to responses at lower $ T $, where the embedding-based similarity calculation captures these semantic discrepancies more effectively.

\begin{figure}[htbp]
    \centering
    \includegraphics[width=0.45\textwidth]{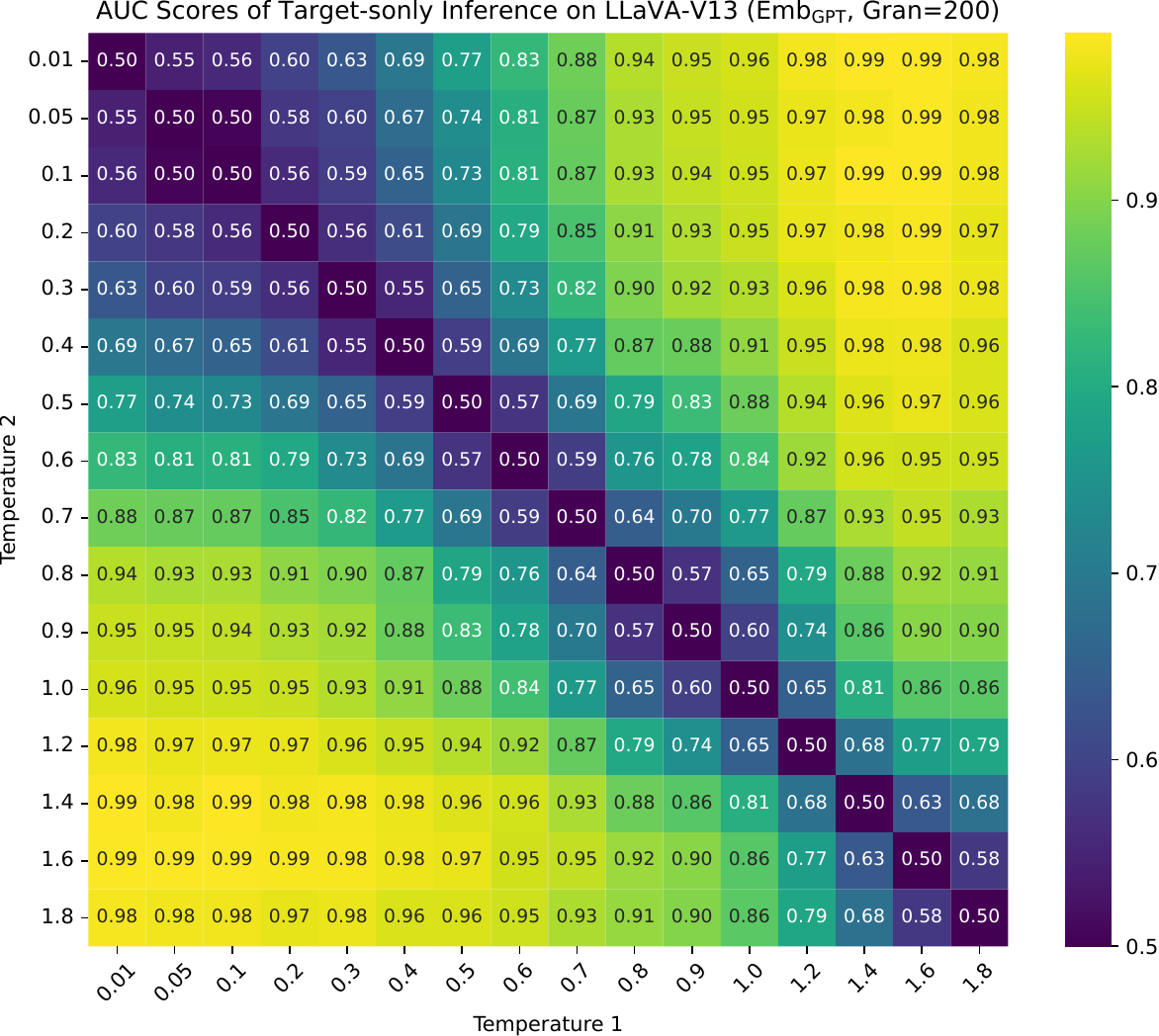}
    \caption{Target-only Inference with Varying $g$ on LLaVA}
    \label{fig:target_only_heat_llava}
\end{figure}

We show the effects of varying $ T_l $ and $ T_h $ on LLaVA with Vicuna-13B in \autoref{fig:target_only_heat_llava}. 
Generally, a larger difference between $ T_l $ and $ T_h $ correlates with higher AUC scores, although larger is not always better. 
Particularly at very high $ T_h $ values, such as 1.8, the quality of model responses is poor, offering limited utility. 
Please refer to \autoref{app:deferred_evaluation} for the results on MiniGPT-4 in \autoref{fig:target_only_heat_minigpt}.

\section{Image-only Inference}
{
In some practical scenarios, the adversary may possess only images without the corresponding text. Therefore, we consider the most challenging scenario in which the adversary has only a set of images that need to be inferred.
}

\subsection{Intuition}
{
Due to inherent randomness in the model generation process, querying the same image multiple times may yield slightly different responses. 
For member sets, since the model has been trained on these sets, the descriptions are more consistent and closely aligned with the ground truth, resulting in greater similarity among them. 
In contrast, for non-member sets, to which the model has not been exposed, the output probability distribution is more uniform, leading to more diverse and less consistent responses. 
Thus, when only images are available, an attacker can repeatedly query the model with the same image, assess the similarity among the responses, and use this information to infer membership status, as illustrated in the bottom line of \autoref{fig:algorithm_overview}.
}

\begin{algorithm}
\caption{{Image-only Inference}}
\label{alg:image_only}
\begin{algorithmic}[1]
\renewcommand{\algorithmicrequire}{ \textbf{Input:}}
\REQUIRE {Target set $\mathbf{X}_v^t$ of size $g$, target model $f_{\theta_t}$, query temperature $T$, threshold $\tau$.}
\FOR{{each $x_v\in\mathbf{X}_v^t$}}
        \STATE {Ask shadow model to describe image $x_v$ $k$ times and obtain  $[r_1, r_2, \cdots, r_k]$}
        \STATE {Compute the similarity score between every pair of these responses and get $[s_1, s_2, \cdots, s_{k\times(k-1)/2}]$}
        \STATE {Average the similarity scores and get $s_{avg}$}
\ENDFOR

\STATE {Compute the mean $\bar{s}_{avg}$}

\IF{{$\bar{s}_{avg} > \tau$}}
    \STATE {Conclude that $\mathds{1} = 1$, i.e., $\mathbf{X}_t$ is a member set}
\ELSE
    \STATE {Conclude that $\mathds{1} = 0$, i.e., $\mathbf{X}_t$ is a non-member set}
\ENDIF
\renewcommand{\algorithmicrequire}{\textbf{Output:}}
\REQUIRE {Membership status $\mathds{1} \in \{0,1\}$}
\end{algorithmic}
\end{algorithm}

\subsection{Methodology}
For each image $ x_i $ within the target set, the target model independently describes the image $ k $ times, yielding $ k $ responses $[r_1, \cdots, r_k]$. The similarity among these $ k $ responses is calculated, resulting in $\frac{k(k-1)}{2}$ similarity scores $[s_1, \ldots, s_{k(k-1)/2}]$, and their mean, $ s_{\text{avg}} $, is calculated. The adversary determines the membership status based on the mean $ s_{\text{avg}} $ of all images in the target set, where a higher $ s_{\text{avg}} $ indicates membership.

\subsection{Evaluation Setting}
This section employs the same target models as described in \autoref{sec:reference_eva_setting}. 
We sample 1,000 sets of granularity $g$ from member data and non-member data, respectively. The images in each set are then fed into \autoref{alg:image_only} for image-only inference.

\subsection{Experimental Results}

\begin{figure}[t]
    \centering
    \includegraphics[width=0.47\textwidth]{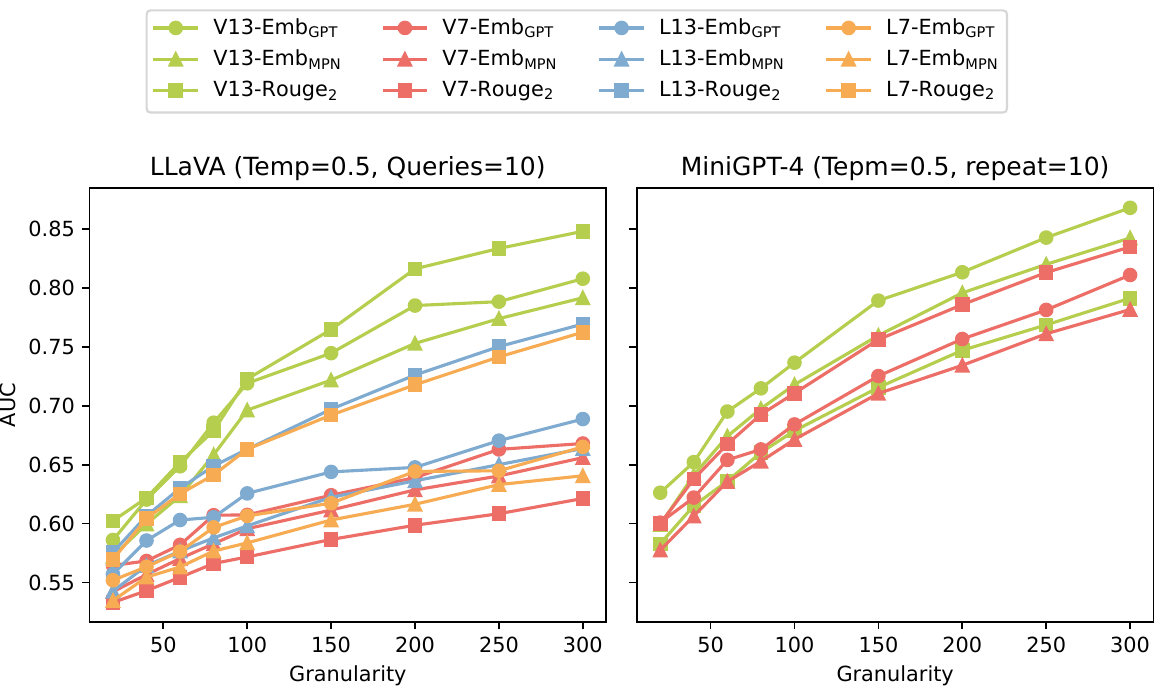}
    \caption{{AUC scores of Image-only Inference}}
    \label{fig:image_only_auc}
\end{figure}

\autoref{fig:image_only_auc} illustrates the impact of granularity, similarity calculation methods, and the type of LLM on the AUC, with the experimental setup specifying $T = 0.5$ and the number of queries $k=10$.
The impact of these variants remains consistent with previous experiments.

\begin{figure}[t]
    \centering
    \includegraphics[width=0.47\textwidth]{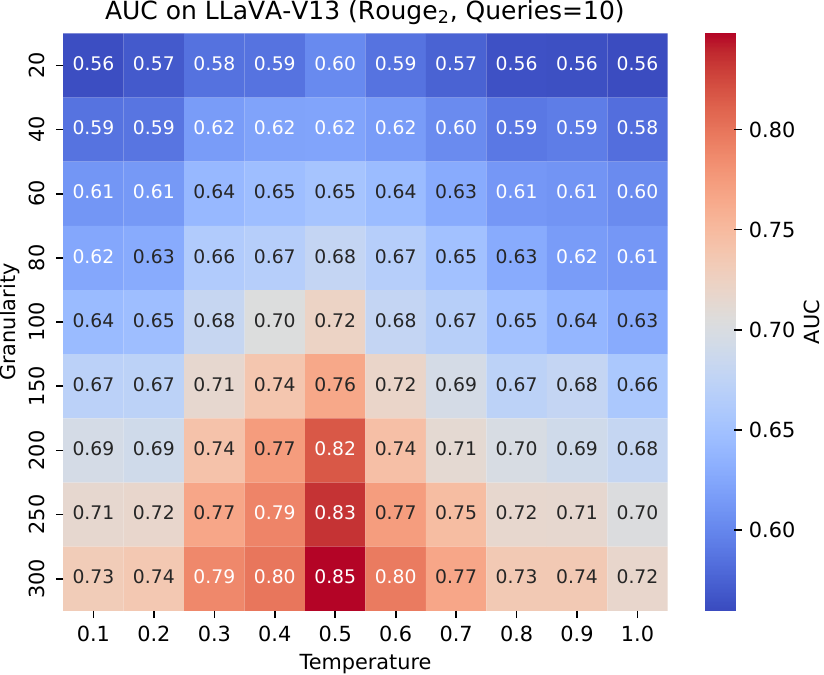}
    \caption{{Image-only Inference with Varing $T$ on LLaVA}}
    \label{fig:image_only_temp}
\end{figure}

\autoref{fig:image_only_temp} demonstrates the effects of temperature on inference success rates. The highest success rate occurs at $T=0.5$. When the temperature is too low, the randomness during model generation is insufficient, leading to very stable descriptions of non-member images by the model, which is inadequate for distinguishing between member and non-member images. Conversely, when the temperature is too high, the excessive randomness results in significant variations between descriptions of member images by the model. Therefore, selecting an appropriate temperature is crucial.

\begin{figure}[t]
    \centering
    \includegraphics[width=0.47\textwidth]{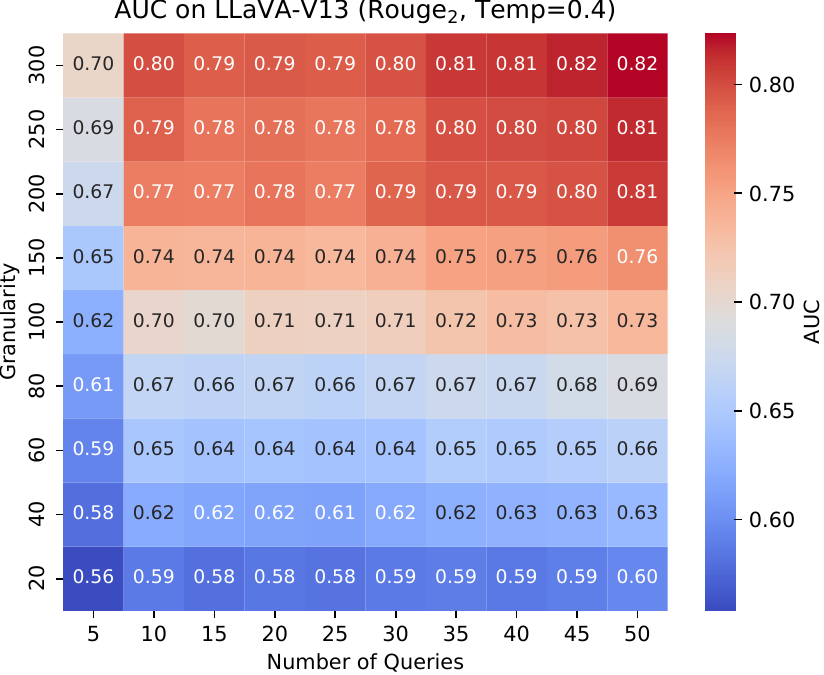}
    \caption{{Image-only Inference with Varing $k$ on LLaVA}}
    \label{fig:image_only_repeat}
\end{figure}

The impact of the number of queries is depicted in \autoref{fig:image_only_repeat}. There is a noticeable improvement when the number of queries increases from 5 to 10; however, further increases in the number of queries only marginally enhance the inference success rate.


\section{Additional Experiments}
This section studies the impact of different settings on the effectiveness of inference methods, including query type, heterogeneous target set, mismatched set size and response length. 
We also discuss and compare the shadow model inference method with existing attacks that share similar threat models.
The experiments in this section are conducted on LLaVA with Vicuna-13B model.
More experimental results can be found in \autoref{app:deferred_evaluation}.

\begin{table*}[htb]
  \caption{\small Comparison of Different Attacks and Query Types on LLaVA-V13.}
  \scriptsize
  \begin{center}
   \setlength{\tabcolsep}{1.7mm}{
    \begin{tabular*}{\textwidth}{ @{\extracolsep{\fill}} l c c c c c c c c c c c c c}
     \toprule
     \multirow{2}{*}{\textbf{Metric}} & \multirow{2}{*}{\textbf{Gran.}} & \multicolumn{3}{c}{\textbf{Shadow Model}} & \multicolumn{3}{c}{\textbf{Reference Member}} & \multicolumn{3}{c}{\textbf{Reference Non-Member}} & \multicolumn{3}{c}{\textbf{Target-only}} \\
     \cmidrule(lr){3-5} \cmidrule(lr){6-8} \cmidrule(lr){9-11} \cmidrule(lr){12-14}
     & & Reason & Detail & Conv. & Reason & Detail & Conv. & Reason & Detail & Conv. & Reason & Detail & Conv. \\
     \midrule
     \multirow{3}{*}{$\text{Emb}_\text{{GPT}}$} & 20 & 88.85\% & 72.72\% & \underline{88.92\%} & \underline{80.89\%} & 69.36\% & 77.67\% & \underline{81.41\%} & 68.23\% & 78.22\% & \underline{78.98\%} & 63.65\% & 74.60\% \\
     & 60 & \underline{97.93\%} & 84.31\% & 97.91\% & 92.19\% & 81.90\% & \underline{92.83\%} & \underline{92.98\%} & 82.20\% & 92.20\% & \underline{89.80\%} & 76.07\% & 88.88\% \\
     & 100 & \underline{99.54\%} & 89.12\% & 99.54\% & 96.87\% & 87.30\% & \underline{96.92\%} & \underline{97.56\%} & 85.73\% & 96.39\% & \underline{94.41\%} & 83.26\% & 93.12\% \\
     \midrule
     \multirow{3}{*}{$\text{Emb}_\text{{MPN}}$} & 20 & 81.97\% & 71.65\% & \underline{82.18\%} & 75.19\% & 65.22\% & \underline{76.58\%} & 73.96\% & 67.98\% & \underline{77.17\%} & \underline{72.35\%} & 63.20\% & 71.79\% \\
     & 60 & \underline{93.76\%} & 82.49\% & 91.99\% & 87.39\% & 78.66\% & \underline{88.19\%} & 87.51\% & 78.73\% & \underline{90.32\%} & 85.09\% & 73.55\% & \underline{85.79\%} \\
     & 100 & \underline{97.30\%} & 88.52\% & 96.89\% & 93.02\% & 82.79\% & \underline{94.81\%} & 93.21\% & 86.13\% & \underline{94.77\%} & \underline{90.51\%} & 81.26\% & 90.08\% \\
     \midrule
     \multirow{3}{*}{$\text{Rouge-2}$} & 20 & \underline{96.94\%} & 86.09\% & 94.47\% & \underline{93.56\%} & 76.41\% & 86.80\% & \underline{93.38\%} & 71.64\% & 87.29\% & 67.44\% & 65.92\% & \underline{69.25\%} \\
     & 60 & \underline{99.87\%} & 97.08\% & 99.49\% & \underline{99.71\%} & 89.47\% & 97.17\% & \underline{99.66\%} & 90.56\% & 97.62\% & 67.43\% & 68.49\% & \underline{83.09\%} \\
     & 100 & \underline{100.00\%} & 99.26\% & 99.88\% & \underline{100.00\%} & 94.51\% & 99.36\% & \underline{100.00\%} & 96.37\% & 99.41\% & 68.94\% & 68.94\% & \underline{87.81\%} \\
     \bottomrule
   \end{tabular*}}
   \end{center}
   \label{tab:llava_data_type}
\end{table*}

\subsection{Query Type}
As discussed in \autoref{sec:shadow_eva_setting}, the instruction tuning dataset for LLaVA is categorized into three types: multi-turn conversations, detailed image descriptions, and complex reasoning. We test the performance of three inference methods across these query types, and the results are presented in \autoref{tab:llava_data_type}. 
Note that the image-only inference is not included as it can only be performed on the type of detailed image description.
Generally, queries of the complex reasoning type achieved the highest success rate in inference, followed by multi-turn conversations, while those of the detailed image description type were less effective. We believe that this is due to the answers in detailed image descriptions being more definitive, allowing the VLM to produce responses very close to the ground truth answer, regardless of whether it has learned the data. In contrast, answers in complex reasoning are more divergent, resulting in larger discrepancies between members and non-members. Similarly, multi-turn conversations, which often involve reasoning questions, also exhibit higher inference success rates.

\begin{figure}[htbp]
    \centering
    \includegraphics[width=0.43\textwidth]{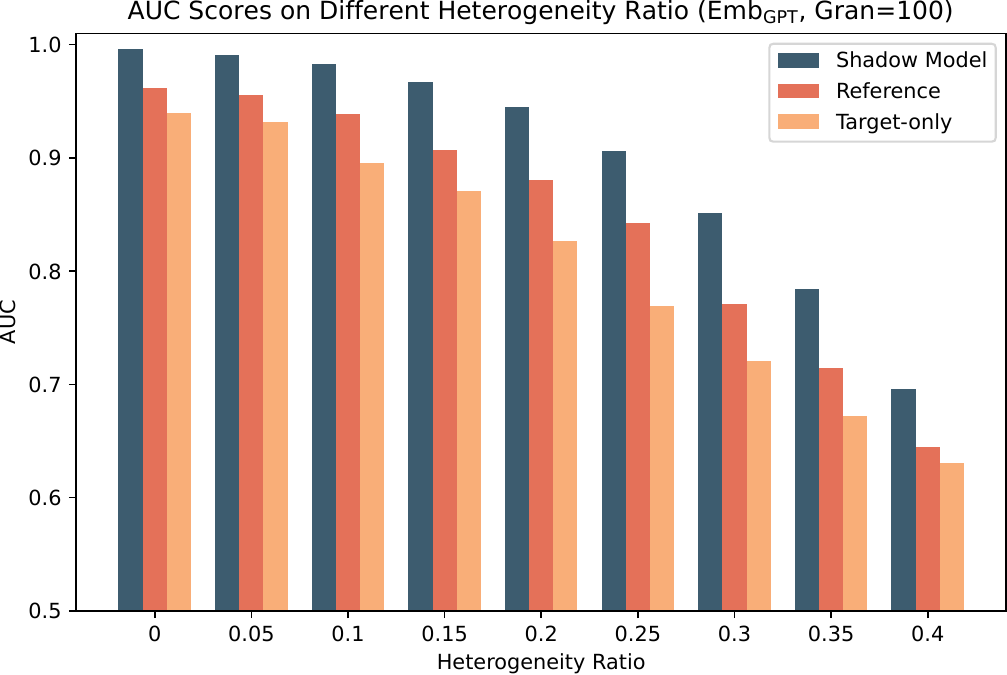}
    \caption{Scores of Heterogeneous Sets}
    \label{fig:heterogeneous}
\end{figure}

\subsection{Heterogeneous Target Set}
\label{sec:heterogeneous}
It is typically assumed that the data samples within a target set share the same membership status. This is because model developers generally aim to comprehensively collect all accessible data, and data in one target set often possess the same accessibility.
For instance, a set of medical images from a single patient is usually either entirely collected or entirely omitted. 

In cases where this uniformity does not hold, we explore the scenario that each target set includes a certain proportion of data with opposing membership statuses. The results, displayed in \autoref{fig:heterogeneous}, show that for all types of inference, the success rate significantly decreases as the heterogeneity ratio increases. However, when the ratio is low (less than 0.2), the inference success rate remains relatively high (higher than 0.8), showing robustness to some extent.

\begin{figure}[htbp]
    \centering
    \includegraphics[width=0.47\textwidth]{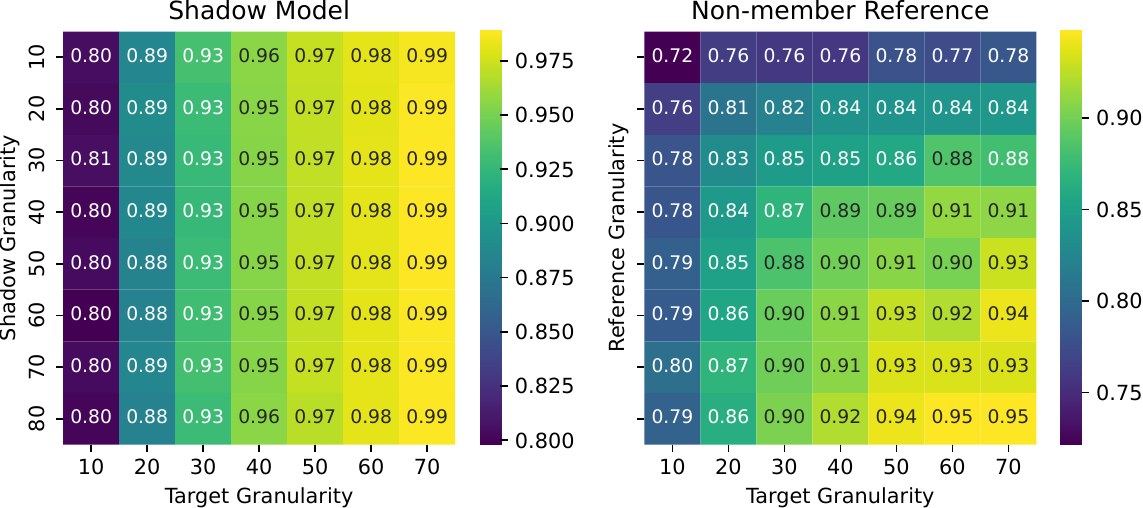}
    \caption{AUC scores of Sets with Mismatched Granularities}
    \label{fig:mismathed}
\end{figure}

\subsection{Mismatched Set Size}
In shadow model inference and reference inference, there are actually two types of granularity: one for the target set and the other for the shadow set or the reference set. We investigate whether a mismatch in the size of these sets affects the success rate of inference and illustrate the experimental results in \autoref{fig:mismathed}. For shadow model inference, the size of the shadow granularity appears to have minimal impact on the success rate of inference. However, the size of the target granularity is positively correlated with the success rate of inference. In the case of reference inference, the success rate is positively correlated with the size of both granularities. Yet, the variation in success rates primarily depends on the size of each granularity itself, rather than on their match, as matched conditions (data along the diagonal) do not significantly outperform adjacent data points.

\begin{figure}[htbp]
    \centering
    \includegraphics[width=0.43\textwidth]{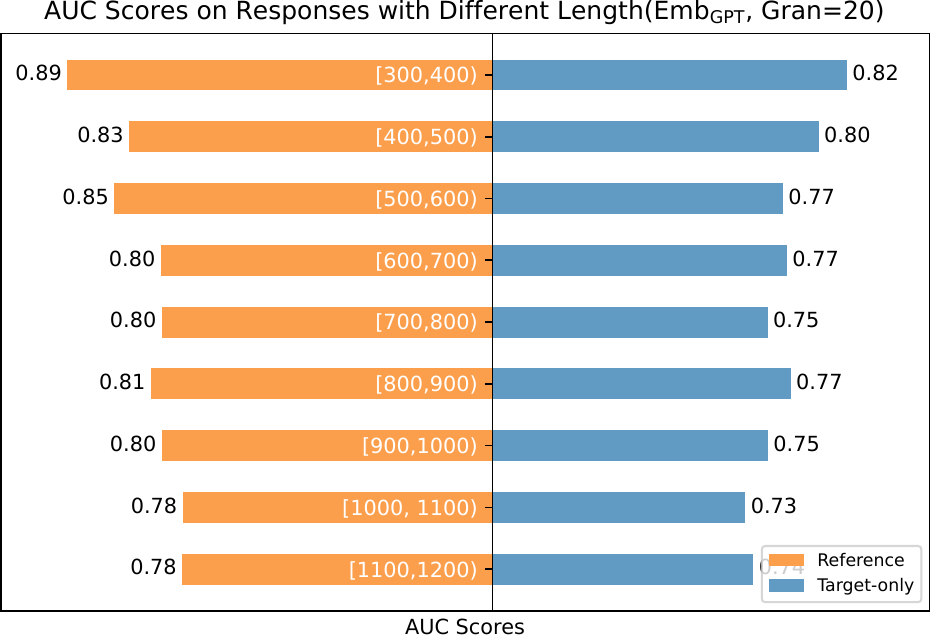}
    \caption{AUC scores on Responses with Different Length}
    \label{fig:response_length}
\end{figure}

\subsection{Response Length}
\label{sec:response_length}
During our experiments, we observe that the similarity between the VLM's response and the ground truth answer correlates with the length of the response for member data. To further investigate, we group all data according to the string length intervals of the responses and conducted inference independently for each group. The results of reference inference and target-only inference, as shown in \autoref{fig:response_length}, indicate that longer responses tend to have lower inference success rates. 

This phenomenon can be attributed to the error accumulation in next-token predictions as described in \autoref{equ:temperature}. Each token's prediction depends not only on the input prompt but also on the tokens already generated in the response. For member data, the initial tokens generated after the image and text prompt are fed into the VLM are primarily determined by the prompt and tend to be closer to the ground truth answer. However, if any token deviates from the answer during generation, this error influences the generation of all subsequent tokens, leading to cumulative deviations from the ground truth answer. As the response lengthens, the performance of member and non-member data converges, increasing the difficulty of differentiation.

Experiment on shadow model inference is not conducted as the data volume varies across different length intervals, which could affect classifier training and result in biased comparisons.

\subsection{Comparison with Existing Work}
\label{sec:m4i_compare}
\begin{figure}[t]
    \centering
    \includegraphics[width=0.43\textwidth]{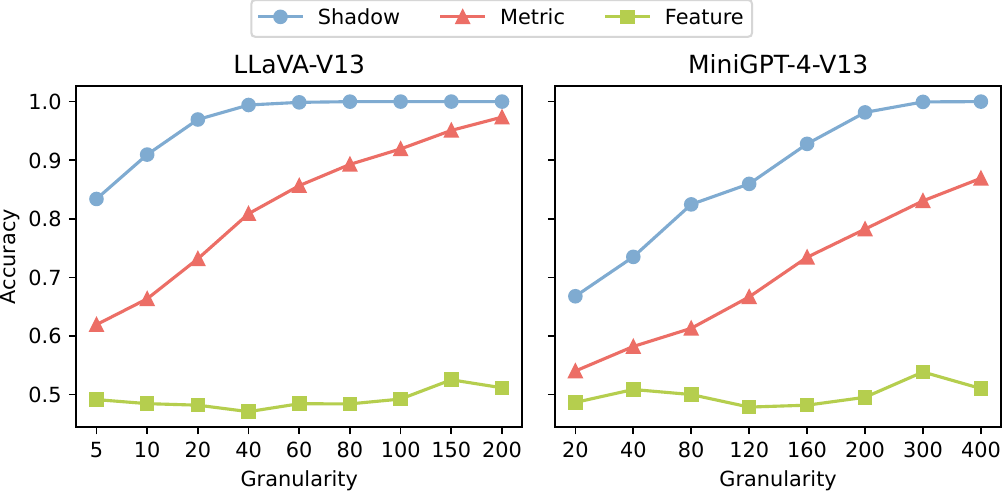}
    \caption{{Comparison of Shadow Model Inference with Exiting MIAs}}
    \label{fig:m4i_comparision}
\end{figure}

While MIAs on LLM-based generative VLMs have not been explored previously, MIAs targeting other multimodal models have been studied. 
For instance, Hu et al.~\cite{hu2022m} introduced two shadow-model-based attacks against a CNN\&RNN-based caption model~\cite{vinyals2015show}. 
This model generates short textual descriptions (around ten words) for input images but lacks the ability to engage in human-like conversation.
Despite significant differences in architecture, data scale, and capabilities between this caption model and modern VLMs, their textual output format makes the proposed MIAs applicable to VLMs. 
Specifically, Their shadow model approach includes two inference methods: (1) a metric-based method that calculates similarity between the model's output and the ground truth using various metrics, and (2) a feature-based method that measures the similarity between images and text in an embedding space, where members typically show higher similarity than non-members. 
Both methods train a classifier on shadow data to leverage these signals for the attack.

As these methods share a similar threat model with our shadow model method, we compare them on LLaVA-Vicuna-13b and MiniGPT-4-Vicuna-13b models. 
To adapt to set-level inference, we use the same sampling policy as before to form sets, and then average the membership signals within every set. 
\autoref{fig:m4i_comparision} shows that under the same threat model, our shadow model inference method significantly outperforms the other methods. The metric-based method’s framework is similar to our method, but its performance is worse due to its non-utilization of temperature-based signals. 
The feature-based method is entirely ineffective, likely because its underlying intuition—that the caption model can semantically align member data pairs well but struggles with non-member data—does not hold for more powerful modern VLMs, which ensure semantic alignment regardless of whether the data was specifically learned.


\section{Discussions}
\label{sec:discussion}
\subsection{Improvements of Proposed Algorithms}
In this paper, we aim to uncover the breadth of threats that set-level membership inference poses to VLMs across various scenarios, rather than optimizing inference algorithms and parameter settings for each specific scenario.
Indeed, the inference algorithms proposed in this paper can be further refined. 
For instance, shadow model inference could benefit from tailored classifier designs for different target models, involving modifications in model architecture, feature selection, and training hyper-parameters. 
As suggested by the results in \autoref{fig:shadow_model_feature}, selecting different feature dimensions could enhance inference performance for varying target models.
Additionally, if a substantial amount of shadow data is available, it would be feasible to train a discriminator directly on the original text, rather than on embeddings. This end-to-end discriminator could capture more detailed information from the original text, avoiding the loss inherent in the embedding process.
For reference inference, we currently query the target model with a single temperature. 
By employing multiple temperatures for querying, the aggregated signal could improve the success rate of the inference. 
A similar enhancement strategy applies to target-only inference, where using multiple pairs of temperatures for querying could boost performance.
Additionally, the experimental results in \autoref{fig:response_length} suggest that focusing on the initial segments of the model response when calculating similarity to the ground truth could more effectively differentiate between members and non-members.

\subsection{Defense Against Membership Inference}

As over-fitting is a fundamental prerequisite for successful membership inference, methods that reduce over-fitting can serve as effective defenses~\cite{hu2023defenses}.
Common strategies such as weight decay (L1/L2 regularization)~\cite{truex2018towards, kaya2020effectiveness}, dropout~\cite{srivastava2014dropout}, and data augmentation are known to decrease over-fitting effectively, albeit often at the expense of model utility. 
Another category of defenses employs differential privacy (DP)~\cite{dwork2006calibrating, abadi2016deep, truex2019effects}, which introduces carefully calibrated noise to provide provable privacy guarantees. However, the introduction of DP noise can significantly compromise model utility, especially for models with a large number of parameters.
Beyond these training-phase defensive measures, machine unlearning~\cite{bourtoule2021machine, guo2020certified, hu2023eraser} seeks to erase the influence of specific data on a trained model. Yet, unlearning algorithms tailored to VLMs remain largely unexplored.

In response to the inference methods proposed in this paper, several basic defenses can be considered, such as restricting users to fixed temperature queries or limiting the range of temperature adjustments. Additionally, as observed in \autoref{sec:response_length}, encouraging longer model responses may help to blur the boundaries between members and non-members.


\section{Related Works}
\subsection{Membership Inference Attack}
Membership inference attacks (MIA) are designed to determine whether a specific data record was used in the training of a machine learning model.
These attacks have been successfully performed across various types of data, including biomedical data~\cite{backes2016membership, homer2008resolving} and mobility traces~\cite{pyrgelis2018knock, pyrgelis2020measuring}. 
Shokri et al.~\cite{shokri2017membership} proposed the first MIA against machine learning models, utilizing multiple shadow models for attack.
Subsequent research~\cite{hui2021practical, li2020membership, yeom2018privacy, salem2019ml, long2017towards, choquette2021label, li2021membership, long2018understanding, ye2022enhanced} progressively relaxed the initial assumptions and introduced more practical approaches, such as data-independent membership inference~\cite{salem2019ml} and decision-based inference~\cite{li2020membership, choquette2021label}, making the inference more practical in realistic scenarios.
While most MIAs focused on image classification models, later studies expanded the application of MIAs to other types of models, including graph models~\cite{hu2023quantifying, he2021stealing}, GANs~\cite{chen2020gan, hilprecht2019monte}, and diffusion models~\cite{hu2021membership, duan2023diffusion}. 

In the context of large language models (LLMs), most privacy attacks have concentrated on extracting private information~\cite{carlini2021extracting, carlini2023quantifying, huang2022large, mireshghallah2022empirical, lukas2023analyzing, nasr2023scalable, hartmann2023sok, liu2024evaluating}. 
Nevertheless, these studies also highlight that privacy extraction only works effectively for worst-case data samples and fails to meet the universality required for MIAs.
Although attempts~\cite{li2023mope, shi2024detecting, mattern2023membership} have been made to adapt MIAs to LLMs, a systematic investigation~\cite{duan2024membership} has shown that existing MIAs for LLMs barely outperform random guessing in most settings across various sizes and domains of LLMs.
This ineffectiveness is largely attributed to the combination of large datasets and few training iterations, a problem that is also faced by large vision-language models.

\subsection{Security and Privacy of VLMs}

While membership inference attacks on VLMs remain largely unexplored, there exists research revolving around other privacy and security issues concerning VLMs.
Xu et al.~\cite{xu2024shadowcast} introduced a poisoning attack designed for VLMs. 
Liang et al.~\cite{liang2024vl} investigated the feasibility of implanting backdoors during the training phase of VLMs. 
Several works~\cite{liu2024arondight, tao2024imgtrojan, qi2024visual} focus on jailbreaking VLMs with adversarial image prompts. 
Additionally, Sun et al.~\cite{sun2024safeguarding} explored adversarial sample attacks and their defenses targeting VLMs.

Prior to the advent of interactive conversation-capable VLMs such as ChatGPT-4~\cite{achiam2023gpt} and LLaVA~\cite{liu2024visual}, VLMs were typically referred to as multi-modal pre-trained representation models such as CLIP~\cite{radford2021learning}, which encode text and images into embeddings within the same embedding space. 
Research on adversarial examples~\cite{yin2024vlattack, zhang2024universal, thota2024demonstration} and membership inference attacks~\cite{ko2023practical, hintersdorf2024does, jayaraman2024d, liu2021encodermi} for such models has been conducted. 
However, due to differences in output modalities, attacks targeting this category of VLMs follow a paradigm significantly distinct from that used for modern generative VLMs, which is the focus of this paper.


\section{Conclusion}
In this paper, we highlight the vulnerabilities associated with membership inference in VLMs by leveraging two observations in overfitting signal capture: the aggregate distribution characteristics of a set of data samples and the differential sensitivity to temperature changes between members and non-members.
By progressively relaxing assumptions about adversarial capabilities, we introduce four membership inference algorithms, thereby demonstrating the extensive privacy threats to VLMs. 
Our comprehensive evaluation confirms the effectiveness of the proposed inference algorithms across various scenarios.

Future work includes: 
1) The inference algorithms could be combined and optimized to adapt to a wider variety of scenarios and enhance inference performance.
2) The inference relies on the assumption of identical distributions between the shadow or reference sets and the target set. The implications of distributional discrepancies need exploration.
3) Further investigation is needed into strategies to effectively mitigate the privacy risks posed by set-level membership inference.
4) Given that temperature is a common parameter in language models, the potential of inference algorithms against LLMs needs further exploration.

\section*{Acknowledgment}
We thank all anonymous reviewers for their constructive comments.
This work is supported by the National Natural Science Foundation of China (No. 62072395), the Joint Fund of the National Natural Science Foundation of China (No. U20A20178), the European Health and Digital Executive Agency (HADEA) within the project ``Understanding the individual host response against Hepatitis D Virus to develop a personalized approach for the management of hepatitis D'' (DSolve, grant agreement number 101057917) and the BMBF with the project ``Repräsentative, synthetische Gesundheitsdaten mit starken Privatsphärengarantien'' (PriSyn, 16KISAO29K).

\bibliographystyle{plain}
\bibliography{reference}

\appendix
\section*{APPENDIX}

\section{Reference Inference with Member Sets}
\label{app:member_reference}
\subsection {Methodology}
While previous implementations of reference inference typically assume a non-member reference set, our inference remains valid if a member set serves as the reference. 
In realistic inference scenarios, large-scale pre-trained models require extensive training data. 
Aside from task-tailored datasets, it is common to utilize all accessible public data. 
Consequently, adversaries could potentially use such public data as member references. 
Moreover, adversaries may employ social engineering techniques to glean information about the training data used, either from publicly available information shared by the model's developers or through private access.
Once the reference data is acquired, the inference is similar to that based on a non-member reference set, with the logic for determining membership status being the inverse, as detailed in \autoref{alg:reference_member}.

\begin{algorithm}
\caption{Reference Inference with Member Set}
\label{alg:reference_member}
\begin{algorithmic}[1]
\renewcommand{\algorithmicrequire}{ \textbf{Input:}}
\REQUIRE Member reference set $\mathbf{X}_r$ of size $g_r$, target set $\mathbf{X}_t$ of size $g_t$, target model $f_{\theta_t}$, threshold $\tau$
\FOR{each $\mathbf{x} = (x_v, x_q, y_a)\in\mathbf{X_r}$}
        \STATE Query shadow model and get $r_r=f_{\theta_t}(x_v, x_q)$
        \STATE Compute similarity score $s_r = sim(r_r, y_a)$
\ENDFOR
\FOR{each $\mathbf{x} = (x_v, x_q, y_a)\in\mathbf{X_t}$}
        \STATE Query shadow model and get $r_t=f_{\theta_t}(x_v, x_q)$
        \STATE Compute similarity score $s_t = sim(r_t, y_a)$
\ENDFOR
\STATE Compute mean $\bar{s}_r/\bar{s}_t$ and standard deviation $\sigma_r/\sigma_t$ of $\mathbf{s}_r/\mathbf{s}_t$
\STATE Calculate the combined standard error $e=\sqrt{\frac{\sigma_t^2}{g_t} + \frac{\sigma_r^2}{g_r}}$
\STATE Calculate the $p$-value $p = 1 - \Phi\left(\frac{\bar{s}_t - \bar{s}_r}{e}\right)$
\IF{$p > \tau$}
    \STATE Conclude that $\mathds{1} = 1$, i.e., $\mathbf{X}_t$ is a member set 
\ELSE
    \STATE Conclude that $\mathds{1} = 0$, i.e., $\mathbf{X}_t$ is a non-member set
\ENDIF
\renewcommand{\algorithmicrequire}{\textbf{Output:}}
\REQUIRE Membership status $\mathds{1} \in \{0,1\}$
\end{algorithmic}
\end{algorithm}

\subsection{Evaluations}

The experimental settings mirror those in \autoref{sec:reference_eva_setting}, except that the reference set is sampled from a member dataset.

\begin{figure}[htbp]
    \centering
    \includegraphics[width=0.47\textwidth]{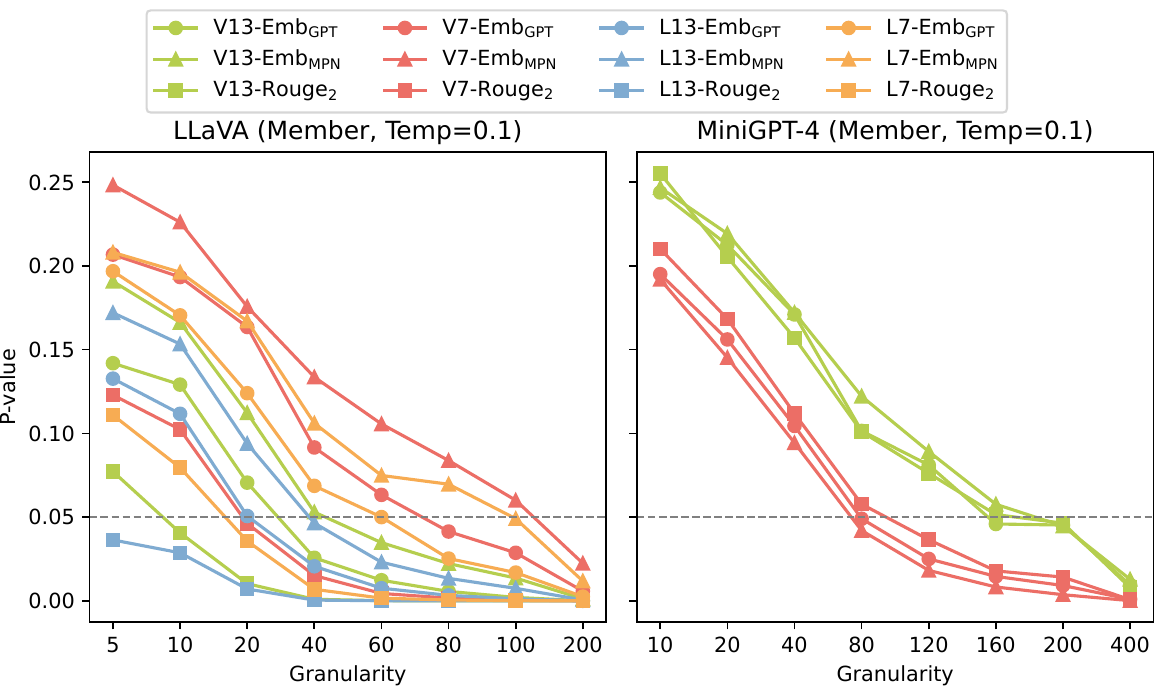}
    \caption{p-Values of Member Reference Inference}
    \label{fig:reference_p_member}
\end{figure}

\autoref{fig:reference_p_member} presents how the $p$-values between reference member sets and target member sets are influenced by granularity, similarity calculation methods, and the type of LLM used. 
The general trends are similar to those in \autoref{sec:reference_eva}: a larger granularity results in lower $p$-values, $p$-values for LLaVA are smaller than those for MiniGPT-4, and the rouge-based calculation outperforms the embedding-based methods. 

\begin{figure}[htbp]
    \centering
    \includegraphics[width=0.47\textwidth]{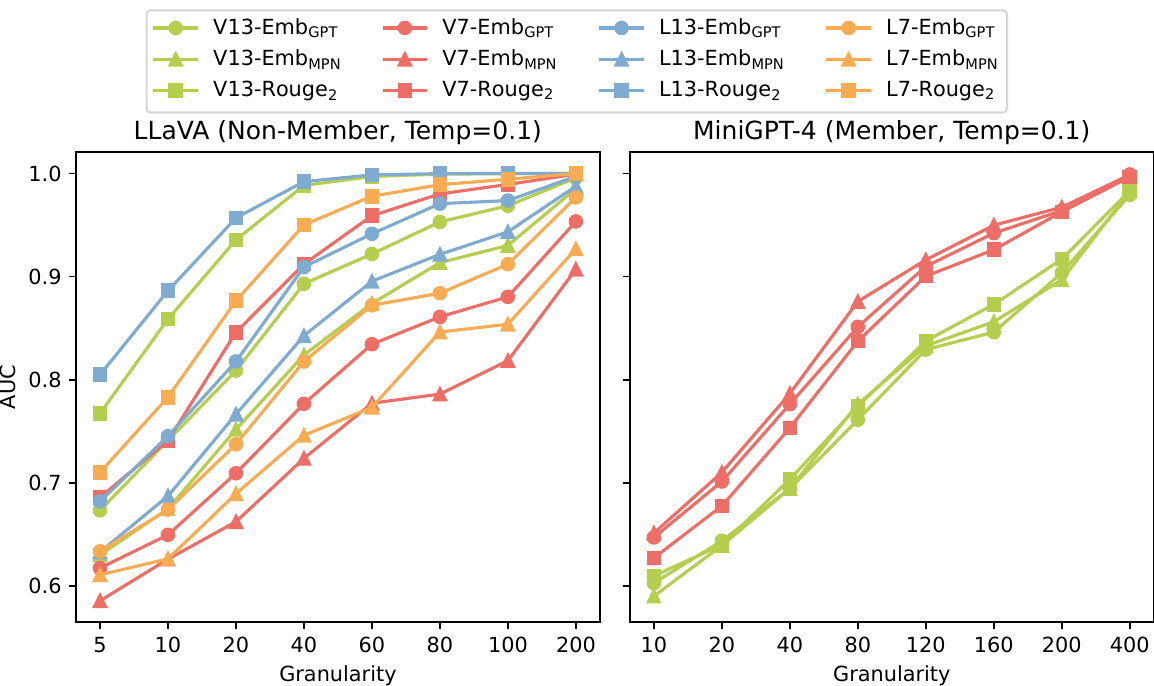}
    \caption{AUC scores of Member Reference Inference}
    \label{fig:reference_auc_member}
\end{figure}

For LLaVA, larger LLM parameters correspond to smaller $p$-values.
The AUC scores in \autoref{fig:reference_auc_member} show trends similar to those in \autoref{fig:reference_p_member}. 

\begin{figure}[htbp]
    \centering
    \includegraphics[width=0.47\textwidth]{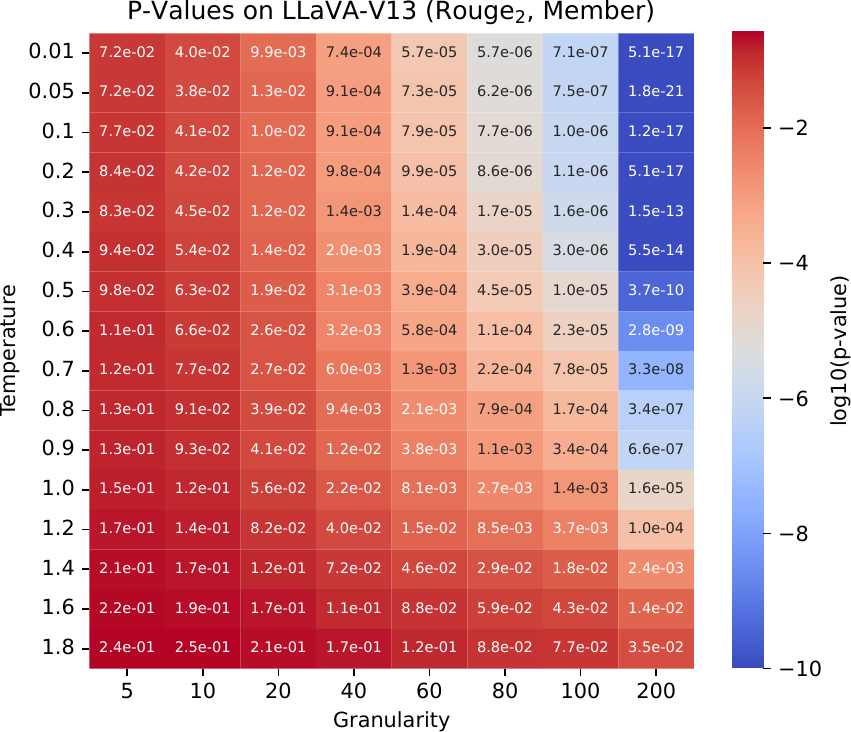}
    \caption{P-Values of Member Reference on LLaVA-V13}
    \label{fig:reference_temp_p_llava_member}
\end{figure}

\begin{figure}[htbp]
    \centering
    \includegraphics[width=0.47\textwidth]{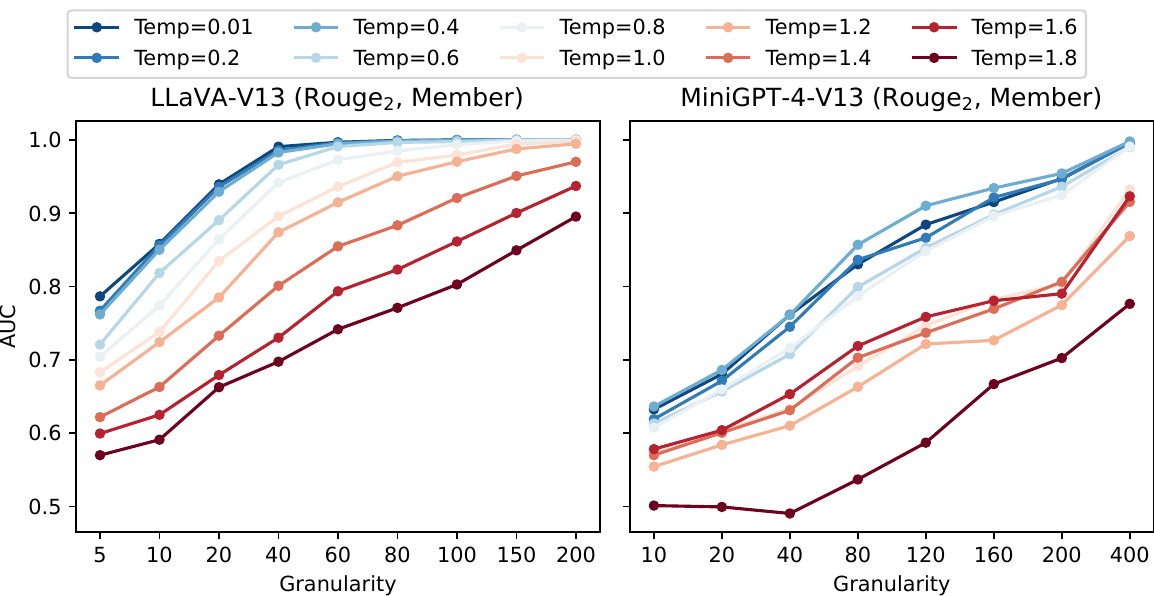}
    \caption{AUC scores of Member Reference Inference}
    \label{fig:reference_temp_auc_member}
\end{figure}

We also display the impact of the effect of varying $ T $ on $p$-values in \autoref{fig:reference_temp_p_llava_member}, and the effect on AUC in \autoref{fig:reference_temp_auc_member}.
The results indicate higher inference success rates with smaller $ T $.

\section {Deferred Experimental Results}
\label{app:deferred_evaluation}
\subsection{Impact of different factors on p-values}
\begin{figure}[htbp]
    \centering
    \includegraphics[width=0.47\textwidth]{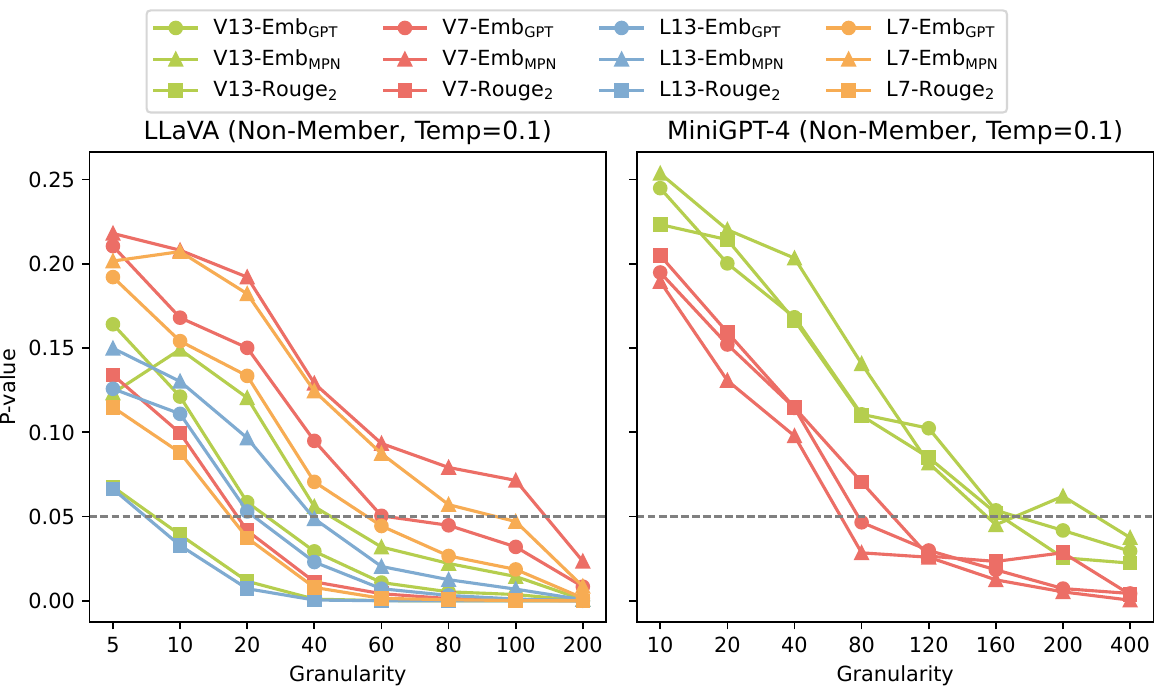}
    \caption{p-Values of Non-member Reference Inference}
    \label{fig:reference_p}
\end{figure}

\autoref{fig:reference_p} presents how the $p$-values are influenced by granularity, similarity calculation methods, and the type of LLM used in reference inference. 
These p-values represent the mean calculated over 1,000 inferences. 
Generally, a larger granularity results in lower $p$-values; $p$-values for LLaVA are smaller than those for MiniGPT-4, and the rouge-based calculation outperforms the embedding-based methods. 
For LLaVA, larger LLM corresponds to smaller $p$-values.

\begin{figure}[htbp]
    \centering
    \includegraphics[width=0.47\textwidth]{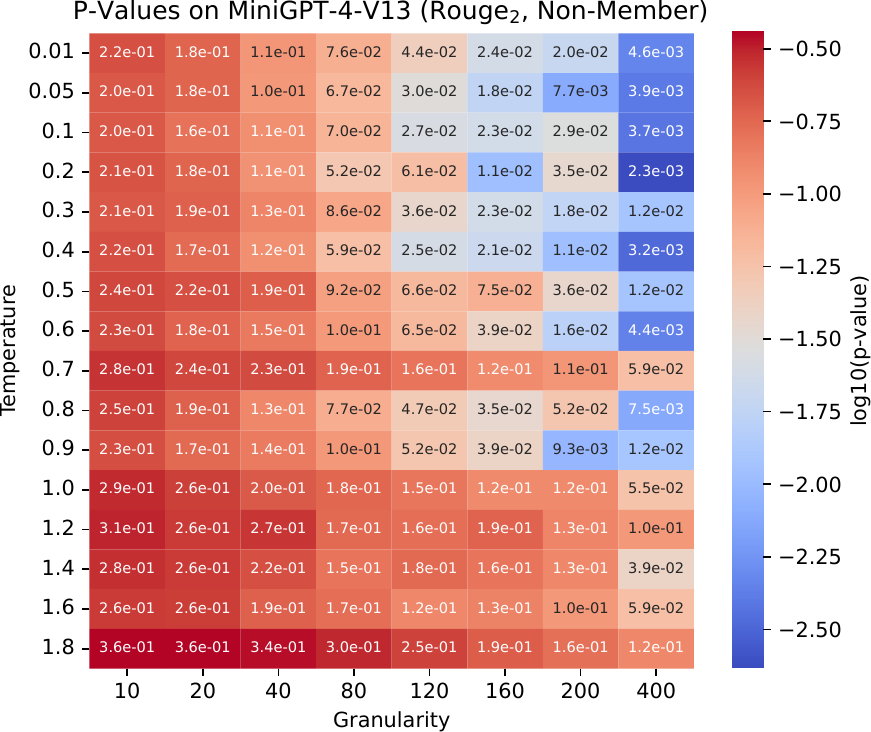}
    \caption{P-Values of Non-member Reference on MiniGPT4}
    \label{fig:reference_temp_p_minigpt}
\end{figure}

\begin{figure}[htbp]
    \centering
    \includegraphics[width=0.47\textwidth]{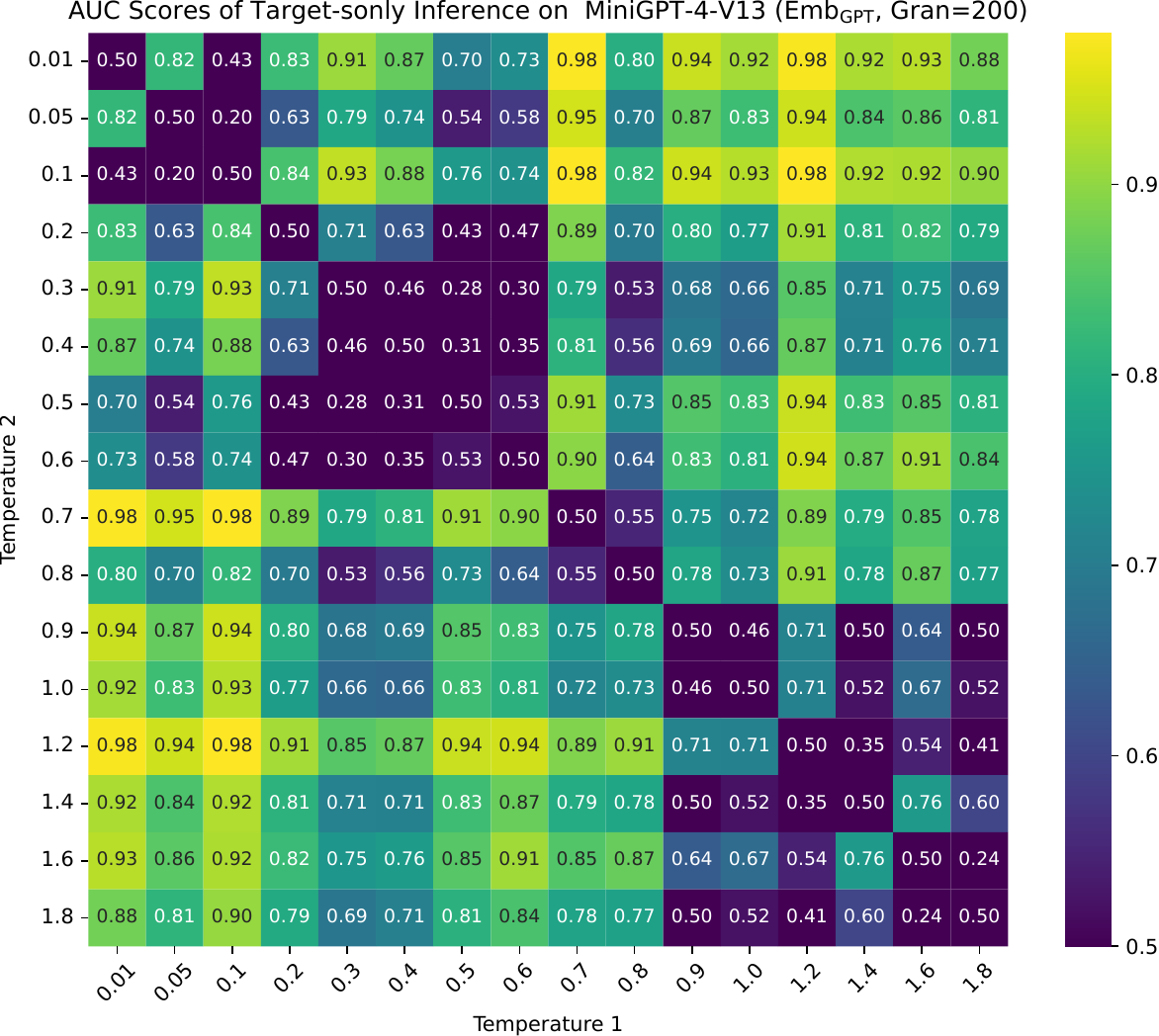}
    \caption{Target-only Inference with Varying $g$ on MiniGPT}
    \label{fig:target_only_heat_minigpt}
\end{figure}

\autoref{fig:reference_temp_p_minigpt} displays the impact of varying $ T $ on $p$-values on MiniGPT-4-V13, and it shows a similar trend as in \autoref{fig:reference_temp_p_llava}.
The effects of varying $ T_l $ and $ T_h $ of target-only inference on MiniGPT-4-V13 are shown in \autoref{fig:target_only_heat_minigpt}. The results align with LLaVA-V13 in \autoref{fig:target_only_heat_llava}.

\end{document}